%% file: main.tex
\documentclass[10pt,conference]{IEEEtran}
\IEEEoverridecommandlockouts

\usepackage{cite}
\usepackage{xurl}
\usepackage{amsmath,amssymb,amsfonts}
\usepackage{algorithmic}
\usepackage{graphicx}
\usepackage{textcomp}
\usepackage{xcolor}

\usepackage{mdframed}

\usepackage{array}
\usepackage{xspace}
\usepackage{diagbox}
\usepackage{booktabs}
\usepackage{enumitem}
\usepackage{hyperref}
\usepackage{multirow}
\usepackage{tcolorbox}
\usepackage{subcaption}

\def\BibTeX{{\rm B\kern-.05em{\sc i\kern-.025em b}\kern-.08em
    T\kern-.1667em\lower.7ex\hbox{E}\kern-.125emX}}
\begin{document}

\newcommand{\sysname}{MATester\xspace}
\newcommand{\rev}[1]{\textcolor{black}{#1}}

\newmdenv[
    backgroundcolor=gray!20,
    linecolor=gray!20,
    linewidth=0pt,
    innerleftmargin=1pt,
    innerrightmargin=1pt,
    settings={\small}
]{finding}

\title{A Comprehensive Study of Implementation Bugs in Multi-modal Agents}

\author{
\IEEEauthorblockN{Suwan Li}
\IEEEauthorblockA{\textit{Department of Computer Science} \\
\textit{Nanjing University}\\
Nanjing, China \\
lisuwan@smail.nju.edu.cn}
\and
\IEEEauthorblockN{Lei Bu}
\IEEEauthorblockA{\textit{Department of Computer Science} \\
\textit{Nanjing University}\\
Nanjing, China}
\and
\IEEEauthorblockN{Shangqing Liu}
\IEEEauthorblockA{\textit{Department of Software Engineering} \\
\textit{Nanjing University}\\
Nanjing, China}
\and
\IEEEauthorblockN{Yile Wang}
\IEEEauthorblockA{\textit{Department of Computer Science} \\
\textit{Nanjing University}\\
Nanjing, China}
\and
\IEEEauthorblockN{Guangdong Bai}
\IEEEauthorblockA{\textit{Department of Computer Science} \\
\textit{The City University of Hongkong}\\
Hongkong, China}
\and
\IEEEauthorblockN{Fuman Xie}
\IEEEauthorblockA{\textit{School of Electrical Engineering and Computer Science} \\
\textit{The University of Queensland}\\
Brisbane, Australia}
\and
\IEEEauthorblockN{Kai Chen}
\IEEEauthorblockA{\textit{Institute of Information Engineering} \\
\textit{Chinese Academy of Science}\\
Beijing, China}
\and
\IEEEauthorblockN{Chang Yue}
\IEEEauthorblockA{\textit{Institute of Information Engineering} \\
\textit{Chinese Academy of Science}\\
Beijing, China}
}

\maketitle

\begin{abstract}
\input{new/abstract}
\end{abstract}


\input{new/intro}

\input{new/mllma}

\input{new/taxonomy}

\input{new/experiment}

\input{new/discussion}

\input{new/related-work}

\input{new/conclusion}


\bibliographystyle{IEEEtran}
\bibliography{reference}

\end{document}

%% file: new/abstract.tex
Multi-Modal Agents (M-agents), empowered by Large Language Models (LLMs), excel in various complex, open-world scenarios such as autonomous driving and robotics. However, their unique \rev{requirements to interact with dynamic and diverse multi-modal environments} introduce novel implementation challenges beyond those faced by traditional agents. \rev{Outdated} perception, untrustworthy planning and \rev{inapplicable} execution could cause traffic accident and financial loss. Despite growing study on agent issues, there has not been a systematic study focusing on M-agent-specific implementation bugs.

To address this gap, we conducted the first systematic study of implementation bugs in M-agents. We collected 34 representative M-agents from diverse sources and, through meticulous filtering, identified 158 M-agent-specific bugs from 1,268 issue reports. Using a top-down strategy, we developed a comprehensive taxonomy that classifies bugs by global symptoms, functionality component-level symptoms, and root causes. We then implemented \sysname, an automatic \rev{proof-of-concept bug identifier} by analyzing runtime inter-component outputs. When applied to 12 extra M-agents, \sysname successfully covered 61.4\% of known open issues and discovered 31 additional bugs, demonstrating the practical usefulness of our study. Our work provides a comprehensive reference and guideline for \rev{classification}, prevention \rev{and fix} of M-agent bugs.

%% file: new/intro.tex
\section{Introduction}

Owing to the rapid advancement of large language models (LLMs), intelligent agents have emerged as a prominent paradigm with capabilities including reasoning \cite{reason_agent_1,reason_agent_2}, program synthesis \cite{SWE_agent_1,SWE_agent_2,SWE_agent_3}, and counseling \cite{consel_agent}. Multi-modal agents (M-agents) further extend this paradigm by interacting with high-dimensional, \textit{real-time} and \textit{heterogeneous} environments, enabling deployment in open-world safety-critical scenarios such as autonomous driving \cite{GPT-Driver,DriveLikeAHuman,GPT4V-AD-Exploration}, robotics \cite{MP5,JARVIS-1,Emma-Alfworld,MC-Planner}, and GUI automation \cite{WebWISE,AutoDroid,MM-Navigator,Auto-GUI}. However, implementation flaws in M-agents' functional components may lead to severe consequences. For instance, outdated environmental perception in autonomous driving can result in traffic accidents \cite{traffic_accident}. Unconstrained execution of unverified plans in GUI automation manifests as unexpected behaviors, potentially leading to financial losses \cite{financial_loss} and privacy violations \cite{privacy_leakage}.

Nevertheless, existing empirical studies primarily focus on single-modal agents, emphasizing aspects such as security \cite{empirical_security}, compliance \cite{empirical_compliance}, code-level implementation defects \cite{agent_fix_agent} \rev{and general module-level issues \cite{llm_agent_module}}. Many works concentrate on specific application domains, like software engineering \cite{understand_SWEA,understand_SWEA_2}, \rev{code generation \cite{understand_codeA}} and search \cite{understand_SA}, or \rev{agent architectures like multi-agents \cite{multiple_agent} and platform-orchestrated agent \cite{platform_orchestrated_agent}}. Although they study bugs from multiple perspectives, there is still a lack of systematic investigation dedicated to M-agents, particularly with respect to \rev{multi-modal environment interaction bugs}.


Compared with single-modal agents, M-agents exhibit three distinctive characteristics: (1) \textit{perception}: M-agents must distill and fuse \textit{real-time} multi-modal information for LLM processing, whereas single-modal agents handle text directly; (2) \textit{execution}: M-agents dynamically adapt action types and parameters to \textit{heterogeneous} environments, while single-modal agents execute relatively fixed actions; (3) \textit{multi-modality}: representations across modalities may introduce cross-modal conflicts. These features substantially complicate M-agents and motivate us to conduct the \textit{first} systematic study of M-agent-specific implementation bugs.


Our study proceeds in three stages. 1) To construct a \textit{representative} dataset, we initially identified 86 M-agents from GitHub, top-tier publications, and surveys. We then performed a coarse-grained text-based filtering by examining associated papers and project documentation, followed by a fine-grained manual code inspection to retain only M-agents with complete functionality components. The manual process is conducted by at least two authors separately to mitigate bias. After that, we collected 1,268 raw reports and manually identified 130 M-agent-specific reports covering 158 distinct bugs.
2) We then classified bugs using a \textit{top-down} strategy across three dimensions: 6 categories of end-user-observable \textit{global symptoms}, 16 categories of developer-oriented \textit{functionality-component-level symptoms}, and 7 categories of \textit{root causes}. Inconsistent perception and inapplicable actions caused by mishandling object dynamism, object informativeness and tool application scenarios are reported in 14.7\% and 50\% of M-agents, respectively. Cross-modal issues span multiple components.
3) Finally, to demonstrate the practical utility of our taxonomy, we implemented \sysname, a preliminary bug detector. \sysname automatically identifies both symptoms by analyzing runtime inter-component outputs. We applied \sysname to 12 extra M-agents, where it successfully uncovered 61.4\% of existing open issues and identified 31 unreported bugs. These results validate the effectiveness of \sysname and highlight the practical value of our proposed taxonomy.

To sum up, this paper has the following three contributions.

\begin{itemize}[leftmargin=*]
    \item To the best of our knowledge, we present the first systematic study of agent-specific implementation bugs in M-agents. Using a step-by-step filtering strategy, we collect 34 M-agents with 1,268 raw issue reports and identify 158 M-agent-specific implementation bugs. \rev{This dataset offers a benchmark for future M-agent testing}.
    \item We propose a taxonomy of M-agent implementation bugs using a top-down analysis strategy. Specifically, the bugs are categorized along three dimensions: global-level symptoms, functionality-component-level symptoms, and root causes. The taxonomy \rev{provides a guideline for understanding, prevention and fix of M-agent bugs}.
    \item We implement \sysname, an automated bug detector for M-agents, to validate our proposed concepts. Based on our findings, \sysname automatically detects symptoms in 12 extra M-agents. It successfully covers 61.4\% of existing open issues and uncovers 31 previously unknown bugs, which demonstrate the applicability of our taxonomy. 
\end{itemize}

%% file: new/mllma.tex
\section{Multi-modal LLM Agents} \label{sec: M-agent}


M-agents are built upon LLMs, leverage external tools to accomplish tasks, and interact with multi-modal environments. Owing to the advanced reasoning, cross-modal processing, and tool-use capabilities, they have been deployed in complex real-world scenarios, including autonomous driving \cite{GPT-Driver,DriveLikeAHuman,GPT4V-AD-Exploration}, robotics \cite{MP5,JARVIS-1,Emma-Alfworld,MC-Planner}, and GUI automation \cite{WebWISE,AutoDroid,MM-Navigator,Auto-GUI}. 

\begin{figure}[!t]
    \centering
    \includegraphics[width=0.9\linewidth]{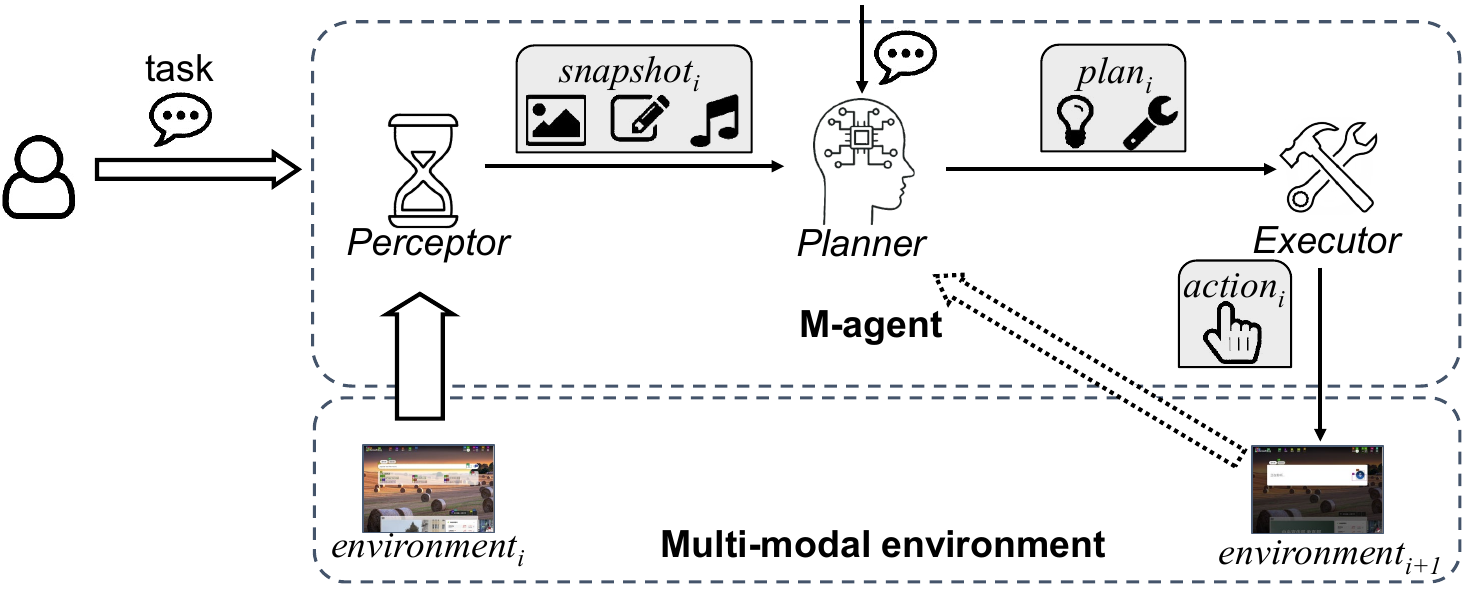}
    \caption{The structure and workflow of M-agents.}
    \label{fig:M-agent}
\end{figure}

Fig.~\ref{fig:M-agent} illustrates the architecture and workflow of a typical M-agent.
\rev{Multi-modal environments often contain dynamic and diverse information that LLMs cannot directly process, so the Perceptor firstly distills it into compact snapshot.} The snapshot is forwarded to the Planner, which generates plans. These plans are subsequently transferred into concrete actions and executed by the Executor. This procedure continues iteratively until the task is complete. \rev{Because the workflow is sequential, failures in any components can propagate downstream, ultimately causing global failure.}


For clarity of presentation, we denote the inter-component outputs as snapshot, plan, and action, using subscripts to distinguish interaction rounds. Specifically, $snapshot_i$ denotes the Perceptor's output in round $i$. The labels $environment_i$ and $environment_{i+1}$ denote the environment state before and after executing $action_i$, respectively.

%% file: new/taxonomy.tex
\section{Study Design} 

\subsection{Overview}
\begin{figure}
    \centering
    \includegraphics[width=0.98\linewidth]{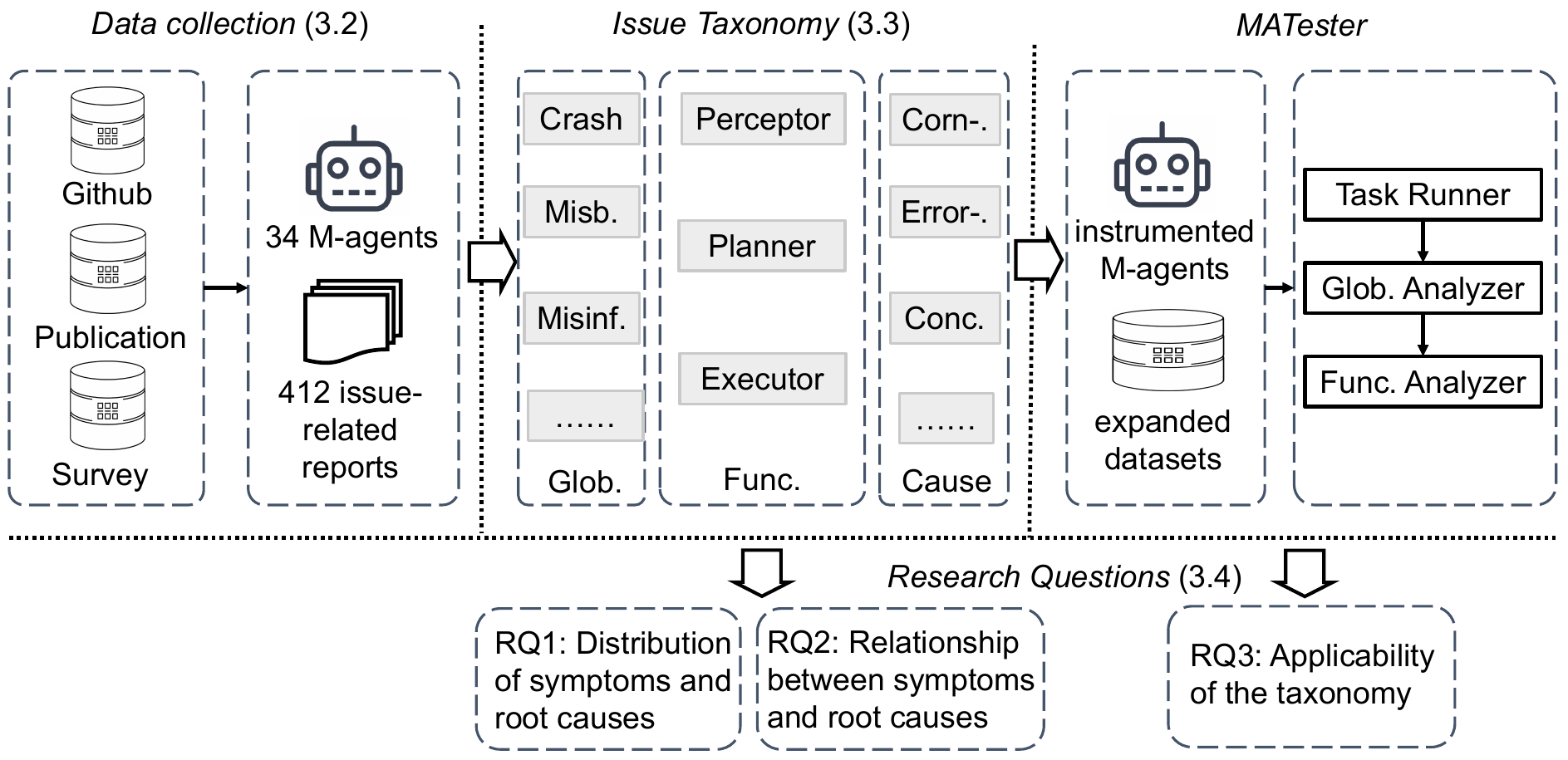}
    \caption{Overview of the analysis.}
    \label{fig: overview}
\end{figure}

The overview of this study is shown in Fig. \ref{fig: overview}. 
To investigate M-agent bugs, we curated an up-to-date collection of M-agents, gathered their issue reports, and focused on agent-specific bugs (Section~\ref{sec: collection}). We constructed a top-down taxonomy analyzing symptoms from a global (end-user) and a functionality-component (developer) perspective, and identified root causes to guide future development (Section~\ref{sec: taxonomy}). Finally, we formulated research questions on the distribution, relationships and applicability of the taxonomy, and implemented \sysname to validate our findings (Section~\ref{sec: rq}).

\subsection{Data Collection and Processing} \label{sec: collection}


To the best of our knowledge, there is currently no comprehensive and up-to-date list of M-agents. Prior to studying bugs, we therefore conducted a systematic search to curate a representative set of M-agents from multiple sources. We then collected issue reports from their code repositories using automated keyword-based searches, followed by rigorous manual inspection and labeling to identify true issue reports.

\subsubsection{M-agents Collection}
To collect M-agents developed in both industrial and academic settings, we gathered candidates from three sources: GitHub repositories, top-tier publications, and recent survey papers. In particular, the survey by \cite{M-LLM-A_survey} provides a comprehensive list of M-agents published prior to February 2024. Given that more than two years have elapsed since the publication of this survey, we further expanded the collection by systematically searching recent publications in leading conferences and journals, including ACL, CHI, CVPR, ICCV, NeurIPS, ICLR, ICSE, FSE, ISSTA, TPAMI, JMLR, TSE and TOSEM, covering the period from 2023 to 2025. All results were collected before December 30, 2025.



Following prior studies \cite{agent_fix_agent,large_language_model_verilog}, we employed the Quasi-Gold Standard (QGS) approach \cite{QGS} via automated keyword searches, snowballing, and manual screening. As shown in Table~\ref{tab: keyword}, keywords on LLMs, multi-modality, and agents were used for initial selection from GitHub and top-tier venues. GitHub repositories with less than 5 stars or forks were excluded. For publications, we applied both backward and forward snowballing \cite{snowballing} to capture overlooked related work.

\begin{table}[!t]
    \centering
    \caption{Keywords used for searching.}
    \label{tab: keyword}
    \vspace{-0.2cm}
    \resizebox{\linewidth}{!}{
    \begin{tabular}{c|c|c|c}
    \toprule
        Type & Large Language Model & Multi-Modal & Agent \\
    \midrule
        Keywords & \shortstack{LLM, Large Language\\Model, Language Model,\\ChatGPT, AI} & \shortstack{Multi-modal, Multi\\modal, Visual, Vision,\\Audio, Speech} & \shortstack{Agent, Embodied,\\Embody, Robot}\\
    \bottomrule
    \end{tabular}
    }
\end{table}


After automated screening, we performed coarse-grained text-based manual filtering by examining topics, descriptions, and README files of GitHub repositories, and titles, keywords, and abstracts of papers, against the criteria in Table~\ref{tab: M-agent-requirement}. Specifically, we excluded: (1) tools that evaluate M-agents rather than implement an M-agent; (2) claimed conversational agents, functioning similarly to LLMs; and (3) agents that accept only text input. This yielded 84 M-agents: 25 from GitHub, 31 from surveys, and 30 from top-tier conferences.

\begin{table}[!t]
    \centering
    \caption{Basic requirements of a M-agent.}
    \label{tab:  M-agent-requirement}
    \vspace{-0.2cm}
    \resizebox{\linewidth}{!}{
    \begin{tabular}{l}
    \toprule
        \textbf{Requirements of a M-agent} \\
    \midrule
        R1: It claims to be implemented on a Large Language Model.\\
        R2: It should be able to process multi-modal environmental information.\\
        R3: It is an agent, at least taking actions to achieve its goal.\\
    \bottomrule
    \end{tabular}
    }
    \vspace{-0.2cm}
\end{table}


Finally, we conducted fine-grained code-based manual filtering to retain runnable M-agents with complete functionality modules. We excluded candidates: (1) lacking an implementation artifact; (2) missing essential files \rev{affecting execution}; (3) relying on unavailable dependencies; (4) lacking core functionality modules; or (5) whose code semantics violate Table~\ref{tab: M-agent-requirement}. This yielded 46 M-agents. Three-fourths (34) of M-agents  were used to construct the taxonomy, while the remaining one-fourth (12) was used to evaluate its extensibility. 

\subsubsection{Bugs Collection}\label{sec: func-issue-collection}

Following prior empirical studies \cite{vehicle_bugs_study,deep_learning_program_study,deep_learning_compiler_study}, we collected closed issues and merged pull requests from selected M-agents' GitHub repositories. To balance issue numbers across repositories with varying activity levels, we applied differentiated strategies. For recently created repositories with fewer than 50 closed reports, or repositories inactive for over a year, we also included open reports. For actively maintained repositories, we collected reports within a 15-month window \cite{deep_learning_compiler_study}. This yielded a raw dataset of 1,268 items.

\rev{To ensure that the issues that we collected are true bugs, we adopted a semi-automatic two-step filtering process.} We first applied an automated filter by searching for bug-relevant keywords, namely ``error'', ``fix'', ``bug'', ``issue'', ``fault'', ``wrong'', ``mistake'' and ``fail'', within issue or pull request titles and tags. This step yielded 263 candidate issues. To complement the automated filtering, a manual inspection of the remaining issues is conducted. Two authors independently reviewed the titles and contents of these issues and reached agreement on 96.5\% of them. Disagreements were subsequently resolved through discussion to reach consensus. Through this process, we identified an additional 149 bug-related issues, resulting in a total of 412 bug-related issues for further analysis.

\subsubsection{Classification and Labeling} \label{sec: labeling}

We classified bugs in two stages. First, we categorized each issue by bug type, then annotated agent-specific implementation bugs with global- and functionality-component-level symptoms and root causes, using prior studies \cite{issue_classifier,issue_discussion_classifier,vehicle_bugs_study,deep_learning_compiler_study,deep_learning_program_study,bug_characteristics} as an initial coding framework. The first author reviewed 11.0\% of issues to validate and refine the taxonomy. All authors then met to consolidate the taxonomy. A second author independently re-labeled the same 11.0\% subset, achieving an inter-rater agreement of 0.42, measured using Cohen's Kappa \cite{Cohen_s_kappa_coefficient}. After resolving disagreements, both authors labeled the remainder independently (Kappa = 0.84), and a final consensus meeting resolved remaining discrepancies. The labeling produced 130 implementation issue reports covering 158 unique bugs, 6 global-level symptoms, 16 functionality-component-level symptoms, and 7 root causes.

\subsection{Bug Taxonomy}\label{sec: taxonomy}
\subsubsection{Global Level Symptoms} \label{sec: global symptom}

We summarized the global-level symptoms, which characterize the observable behaviors of M-agents from an end-user perspective when bugs occur. The identified global-level symptoms are described below.

\begin{itemize}[leftmargin=*]
    \item \textit{Report error and crash \rev{(Crash)}.} This symptom characterizes the unexpected termination of M-agents during task execution, typically accompanied by error reports.
    \item \textit{Misbehave during execution without crash \rev{(Misbehave)}.} This symptom describes incorrect agent behavior without causing a system crash. Specifically, the agent performs actions that are irrelevant to the given task or inapplicable to the environment, ultimately resulting in task failure.
    \item \textit{Can not respond \rev{(Unrespond)}.} This symptom indicates that M-agents become unresponsive for an extended period.
    \item \textit{Mis-notify the exit \rev{(Mis-signal)}.} This symptom characterizes the proactive exit of M-agents before the task completion.
    \item \textit{Behave inconsistently under the same setting \rev{(Diverge)}.} This symptom characterizes inconsistent behavior, whereby an M-agent exhibits different execution outcomes when performing the same task under identical environmental conditions. Notably, this is a result-oriented rather than an action-oriented symptom: if an M-agent follows different actions to complete the same task but lead to identical outcomes, the behavior is not considered inconsistent.
    \item \textit{User interface provides wrong information \rev{(Misinform)}.} This symptom denotes inconsistency between the M-agent’s behavior and the feedback it provides to users.
\end{itemize}


\subsubsection{Functionality Component Level Symptoms} \label{sec: functionality symptom}

Since an M-agent comprises three core functionality components, functionality-component-level symptoms are correspondingly categorized according to these components.


\noindent\textbf{Perceptor}.
The Perceptor transforms the multi-modal environment information into a snapshot for the Planner's LLM. When it fails, the following symptoms arise.

\begin{itemize}[leftmargin=*]
    \item \textit{The snapshot does not exist \rev{(Missing-S)}.}
    \item \textit{The snapshot has wrong format \rev{(Malformed-S)}.} \rev{M-agents' snapshots are typically multi-modal or non-textual. This inherent complexity makes them more susceptible to formatting issues. Moreover, non-textual snapshots are typically stored as files. Therefore, bugs related to file names and file paths may also lead to this symptom.}
    \item \textit{The snapshot's content is wrong or incomplete compared with the environment \rev{(Wrong-S)}.} \rev{The environmental heterogeneity may affect the Perceptor's ability to recognize specific objects (incomplete), while the dynamic feature may cause outdated snapshots (wrong).} We further split it into two sub-symptoms ``same (SM) / different (DM)'' according to modality consistency of snapshots and environments. 
    \item \textit{Available objects are labeled in the snapshot, but labels are wrong or incomplete \rev{(Mislabeled-S)}.} \rev{Due to the diversity of objects in multi-modal environments, Perceptors normally label accessible objects to assist the LLMs in identifying them. However, these labels may be inaccurate: accessible objects are missed (incomplete), or environmental features, like the background, are marked (wrong).}
    \item \textit{Two environments are the same, but their snapshots are different \rev{(Inconsistent-S)}.} \rev{Environments are the same if they contain the same objects with the same position, outlook and available actions. Due to the dynamic environment and instable agent behavior, the Perceptor may generate different snapshots for the same environment.} We further split it into two sub-symptoms ``same (SM) / different (DM)'' according to modality consistency of snapshots and environments.
\end{itemize}

\noindent\textbf{Planner}.
The Planner constructs a prompt from the snapshot and the task, then queries the LLM for the next plan. Bugs may arise in both prompt construction and plan generation.

\begin{itemize}[leftmargin=*]
    \item \textit{The prompt misses necessary information \rev{(Wrong-PR)}.} A complete prompt should include at least the following information: environmental information (the snapshot), task specification, available tools and expected output formats.
    \item \textit{The plan does not exist \rev{(Missing-P)}.}
    \item \textit{The plan has wrong format \rev{(Malformed-P)}.} The plan should satisfy the output format specified in the prompt.
    \item \textit{The plan is internally contradictory \rev{(Conflict-P)}.} It mainly exhibits two types of forms. The LLM may ``act'' (i.e., invoke available tools) differently from it ``thinks'' (i.e., expresses its thought in natural language). Tool invocations may also exhibit parameter or behavior conflicts. 
    \item \textit{The plan does not satisfy constraints given by the snapshot \rev{(Inapplicable-P)}}. \rev{The snapshots are often multi-modal or non-textual, which brings challenges to object identification and action extraction.} Therefore, LLM-generated plans may contain wrong actions or unsupported parameters. We further split it into sub-symptoms ``same (SM) /different (DM)'' based on modality consistency of plans and snapshots.
    \item \textit{The plan is not related to the task \rev{(Irrelevant-P)}.} The plan should facilitate task completion. We further classify it into two types, representing ``when'' \rev{(Wrong-Stop)} and ``what'' \rev{(Wrong-Step)} decisions respectively.
    \begin{itemize}[label=-]
        \item The plan cannot correctly decide the task termination
        \item The plan cannot lead to task completion
    \end{itemize}
    \item \textit{The tasks, snapshots and prompts are nearly the same, but their plans are different and finally result in different execution results \rev{(Inconsistent-P)}.} \rev{Multi-modal inputs could amplify the inherent uncertainty of LLMs \cite{multi-modal-hallucination}, making them likely to produce different plans given the same input.} However, the randomness of the plan should not affect the execution results. This symptom is classified into sub-symptoms ``same (SM) /different (DM)'' based on modality consistency of plans and snapshots.
\end{itemize}

\noindent\textbf{Executor}.
The Executor component translates textual plans into concrete actions by invoking off-the-shelf or custom-defined APIs to advance task execution. \rev{It is the Executor's responsibility to decide when and how to interact with the multi-modal and dynamic environment.} When the Executor encounters bugs, they manifest as the following symptoms.

\begin{itemize}[leftmargin=*]
    \item \textit{The action does not exist \rev{(Missing-A)}.} This symptom is identified if M-agents do nothing following correct plans. 
    \item \textit{The action is different from the plan \rev{(Wrong-A)}.} Given a correct plan, the action should follow its instruction.
    \item \textit{The action is not applicable to the environment \rev{(Inapplicable-A)}.} \rev{Since the environment is constantly changing, plans made on outdated snapshots may not be applicable to the current environment.} It is identified if the action follows the plan's instruction, but is not applicable to the environment. It is classified into two types.
    \begin{itemize}[label=-]
        \item The plan is wrong, the action sticks to the plan and causes crashes \rev{(Wrong-P\&A)}. Normally, an action following a wrong plan is not identified as an Executor-related bug unless the M-agent crashes, because only the wrong action is the direct cause of Crash.
        \item The plan and the invoked tool are correct, but the tool runs unexpectedly \rev{(Wrong-API)}.
    \end{itemize}
    \item \textit{The reflection on the action's result is wrong \rev{(Wrong-R)}.} This symptom is identified if the reflection of the action's influence on the environment is different from the fact.
\end{itemize}

\subsubsection{Root Cause}
Root causes constitute another category of M-agent analysis in this work. \rev{Below, we only introduce the major categories that are independent of the scenario. The specific root causes will be introduced in Section \ref{sec: experiment}.}

\begin{itemize}[leftmargin=*]
    \item \textit{Concurrency}. It refers to situations in which multiple components access shared resources without proper synchronization. We observed two representative scenarios. The first is a data race, where the environment undergoes internal changes (i.e., a write to the environment) after $snapshot_i$ is captured (i.e., a read of the environment), rendering $action_i$ inapplicable to the updated environment. In this case, no happens-before relationship exists between the environment write and the snapshot read. The second scenario is an atomicity violation, in which $snapshot_{i+1}$ is captured after $action_i$ is executed, but the environment has not yet transitioned to a stable state $environment_{i+1}$ under the effect of $action_i$. Consequently, $snapshot_{i+1}$ does not accurately represent $environment_{i+1}$, thereby adversely affecting subsequent stages of the agent’s workflow.

    \item \textit{\rev{Corner-case gaps}}. It arises when functionality modules fail to correctly handle edge cases, which typically stem from rare yet valid inputs. \rev{Based on the locations of corner cases, we further categorize the overlooked factors into \textit{task processing}, \textit{plan comprehension}, \textit{snapshot generation}, \textit{LLM query}, and \textit{tool understanding and invocation}.
    For example, the Perceptor may ignore ``dynamic objects'', and the Executor may invoke tools with limited ``applicable scenarios'' and operate on unauthorized objects. 
    The detailed classification is introduced and analyzed in Section \ref{sec: experiment}}.

    \item \textit{\rev{Error-case gaps}}. It refers to situations in which functionality modules lack robustness to minor errors, allowing small deviations to propagate into more severe failures. \rev{Similar to corner-case gaps, the overlooked factors are also divided under different phases: \textit{plan comprehension}, \textit{snapshot generation}, \textit{LLM query}, and \textit{tool understanding and invocation}. For example, facing an ``unexpected environment'', the invoked tool may access unavailable objects or perform unsupported actions, causing Misbehave or Crash. The detailed classification is presented in Section \ref{sec: experiment}}.

    \item \textit{\rev{Lack of persistent memories for task execution experience (No persistent memory)}}. It captures bugs arising from the absence of a memory mechanism in M-agents \rev{that conduct user-specified tasks}, which adversely affects behavioral consistency. When equipped with memory, M-agents can leverage prior successes and failures to inform subsequent decision making. In contrast, \rev{a task-oriented} M-agent treats repeated instances of the same task as entirely new, independent executions. Moreover, LLMs inherently exhibit a degree of stochasticity, so the lack of memory further affects the reproduction of previously successful behaviors.

    \item \textit{LLM's limitation}. LLMs exhibit inherent limitations, including hallucination, inconsistency, and instability. In M-LLMs, such inconsistency may manifest across different modalities.

    \item \textit{Unsuitable LLM parameters (\rev{Bad parameters})}. The behavior of the underlying LLM is highly sensitive to configuration parameters, including \textit{prompt design}, \textit{temperature setting}, and \textit{maximum token limits}. Prompts that omit critical information or exceed model constraints can significantly impair plan generation. An inappropriate temperature setting may lead to overly rigid or constraint-violating behaviors, while an insufficient maximum token limit can result in truncated or incomplete outputs.

    \item \textit{Incorrect code semantics (\rev{Incorrect semantics})}. 
    We observed two representative forms of such errors. Variables associated with the $i$-th interaction round are \textit{incorrectly updated} for the $(i+1)$-th round. In anther case, the implemented code semantics \textit{deviate substantially} from the intended behavior implied by comments or function names. This root cause does not follow a systematic pattern and is likely attributable to careless or erroneous implementations.
\end{itemize}

\vspace{-0.2cm}
\subsection{Research Questions} \label{sec: rq}

Our study aims to answer the following research questions.




\noindent\textbf{RQ1: What is the proportion and the distributions of agent-specific implementation bugs?} It characterizes the reliability landscape of M-agents by quantifying how frequently agent-specific bugs occur and how they are distributed, revealing that such problems are systematic rather than isolated.

\noindent\textbf{RQ2: What is the relationship between these symptoms?} It analyzes how agent-specific bugs manifest and propagate across functionality components, revealing symptom co-occurrence and causal relationships to enable more effective diagnosis and repairing strategies.

\noindent\textbf{RQ3: Is our taxonomy applicable to undiscovered bugs?} It evaluates whether the taxonomy generalizes to undiscovered bugs, validating it as a predictive and diagnostic tool rather than a purely descriptive framework.

%% file: new/experiment.tex
\section{Experimental Results} \label{sec: experiment}

\subsection{RQ1: Distribution}
We collected 412 reports and classified them into various bug types. The agent-specific implementation bugs are further analyzed for global symptoms, functionality component-level symptoms, and root causes.

\noindent\textbf{Distribution of bug types.}
Table \ref{tab: issue type} shows the distribution of bug types. 
Implementation bugs account for the largest proportion of about 38.3\%.
They are further classified into agent-specific and irrelevant bugs. Irrelevant bugs refer to bugs of assistive functionalities, such as dataset processing and log management. We observe that the number of issues containing agent-specific implementation bugs is 4.6 times that of irrelevant implementation bugs. The first column also reports that there are 158 bugs found in agent-specific implementation issue reports.


Dependency, configuration, version, and other non-implementation-bugs account for the remaining categories and are not directly linked to M-agent behavior. Symptoms and root causes are thus identified solely from agent-specific implementation bugs.

\begin{table}[!t]
    \centering
    \caption{Distribution of bug types. The first type is reported in the format of ``issue number / bug number''.}
    \label{tab: issue type}
    \vspace{-0.2cm}
    \resizebox{\linewidth}{!}{
    \begin{tabular}{c|c|c|c|c|c|c|c|c}
    \toprule
        \shortstack{Impl.\\bug\\(specif.)} & \shortstack{Impl.\\bug\\(irrel.)} & \shortstack{Depend\\-ency\\bug} & \shortstack{Cross\\-system\\bug} &  \shortstack{Config.\\bug} & \shortstack{Usage\\issue} & \shortstack{Version\\bug} & \shortstack{Import\\bug} & \shortstack{Not\\a\\bug} \\
    \midrule
        \textbf{130} / 158 & 28 & 57 & 28 & 45 & 12 & 24 & 25 & 63\\
    \bottomrule
    \end{tabular}
    }
    \vspace{-0.2cm}
\end{table}

\begin{finding}
    \textbf{FINDING 1}: \textbf{Agent-specific implementation bugs} account for the \textbf{largest proportion} among all the bug types. They are directly related to M-agent's behavior, and potentially reveal the issues in its functionality components.
\end{finding}

\noindent\textbf{Distribution of global level symptoms of agent-specific implementation bugs.}
Table \ref{tab: symptom} shows the distribution of global level symptoms. Crash is the most common, accounting for over half of all bugs. Misbehave follows at 36.7\%. Runtime symptoms (Unrespond, Mis-signal) and functionality symptoms (Diverge, Misinform) are also observed.

\begin{table}[!t]
    \centering
    \caption{Distribution of global level symptoms and percentage of M-agents with these symptoms.}
    \label{tab: symptom}
    \vspace{-0.2cm}
    \resizebox{0.85\linewidth}{!}{
    \begin{tabular}{c|cccccc}
    \toprule
        Glob. & \rev{Crash} & \rev{Misb.} & \rev{Unresp.} & \rev{Mis-sig.} & \rev{Diverge} & \rev{Misinf.} \\
    \midrule
        Num. & \textbf{83} & \textbf{58} & 5 & 3 & 5 & 4\\
    \midrule
        Perc. & \textbf{38.2\%} & \textbf{35.3\%} & 8.8\% & 8.8\% & 8.8\% & 8.8\%\\
    \bottomrule
    \end{tabular}
    }
    \vspace{-0.5cm}
\end{table}


\rev{Compared with single-modal agents, M-agents are more deployed in real-life scenarios, causing more severe consequences.} Crash directly interrupts tasks and is observed in autonomous driving, GUI automation, and robotics, leading to outcomes like traffic accidents. \rev{Misbehave can cause financial leakage or privacy violations (e.g., inserting wrong account IDs or phone numbers into text boxes).} Diverge and Misinform indicate unstable or untrustworthy agent behavior.

\begin{finding}
    \textbf{FINDING 2}: \textbf{Crash} and \textbf{Misbehave} are the most common global symptoms, accounting for \textbf{52.5\%} and \textbf{36.7\%} of M-agent-specific bugs, respectively. 
    Crash causes traffic accidents in autonomous driving. Misbehave manifests as unexpected tokens inserted into text boxes of mobile apps, further causing financial or privacy leakage.
\end{finding}

\noindent\textbf{Distribution of functionality component level symptoms in agent-specific implementation bugs.}
Table \ref{tab: functionality_component_symptom} shows the distribution of functionality component level symptoms classified by different functionality components, along with the percentage of M-agents with specific symptoms.

\rev{The multifacetedness, complexity and dynamism of the multi-modal environments pose challenges to acquire complete, accurate and real-time snapshots and identify interactive objects, resulting in the prevalence of Wrong-S and Mislabeled-S.}
Wrong-S is discovered in 9 cases and revealed in 14.7\% of M-agents. Mislabeled-S is found in 10 cases.

\rev{Among the two subcategories of Wrong-S, DM is supported by more cases than SM because cross-modality environmental perception is more challenging. In addition, even SM perception in M-agents is more prone to errors due to the involvement of non-textual modality object identification and understanding.
Identifying and labeling objects in M-agents are necessary especially when the environments contain complex backgrounds or distracting elements, making Mislabeled-S a unique M-agent symptom. Its prevalence stems from M-agents' failure to recognize \textit{dynamically generated objects}, which are common in multi-modal environments. M-agents that percept based solely on static layouts often miss dynamically generated objects.
}

\begin{finding}
    \textbf{FINDING 3-1}: \textbf{Wrong-S} is the most severe \textbf{Perceptor} symptom.
    Its two subcategories are evenly distributed, indicating that both \textbf{cross-modality and non-textual intra-modality environment perception} are challenging. Due to the diversity of the multi-modal environments, \rev{\textbf{Wrong-S} is also common as Perceptors could overlook \textbf{specific object types}.}
\end{finding}

\begin{table}[t]
    \centering
    \caption{Distribution of functionality component level symptoms and percentage of M-agents with these symptoms. The values in Wrong-S, Inconsistent-S, Inapplicable-P and Inconsistent-P are in the form of total(SM, DM). The values in Irrelevant-P and Inapplicable-A are total(Wrong-Stop, Wrong-Stop) and total(Wrong-P\&A, Wrong-API), respectively.}
    \label{tab: functionality_component_symptom}
    
    \begin{minipage}{\linewidth}
        \centering
        
        \resizebox{0.8\linewidth}{!}{
            \begin{tabular}{c|ccccc}
            \toprule
                Symp. & \multicolumn{5}{c}{Perceptor}\\
            \cmidrule{2-6}
                & Missing-S & Malformed-S & Wrong-S & Mislabeled-S & Inconsistent-S\\
            \midrule
                Num. & 6 & 8 & \shortstack{\textbf{9}\\(4, 5)} & \textbf{10} & \shortstack{4\\(2, 2)}\\
            \midrule
                Perc. & 8.8\% & 8.8\% & \textbf{14.7\%} & 5.9\% & 5.9\%\\ 
            \bottomrule
            \end{tabular}
        }
    \end{minipage}
    
    \vspace{0.2cm}
    
    \begin{minipage}{\linewidth}
        \centering
        \resizebox{0.98\linewidth}{!}{
            \begin{tabular}{c|ccccccc}
            \toprule
                Symp. & \multicolumn{7}{c}{Planner}\\
            \cmidrule{2-8}
                & Wrong-PR & Missing-P & Malformed-P & Conflict-P & Inapplicable-P & Irrelevant-P & Inconsistent-P\\
            \midrule
                Num. & 5 & \textbf{15} & \textbf{15} & 2 & \shortstack{\textbf{14}\\(9, 5)} & \shortstack{\textbf{14}\\(3, 11)} & \shortstack{3\\(2, 1)}\\
            \midrule
                Perc. & 2.9\% & 14.7\% & 11.8\% & 2.9\% & \textbf{23.5\%} & \textbf{26.5\%} & 8.8\%\\ 
            \bottomrule
            \end{tabular}
        }
    \end{minipage}
    
    \vspace{0.2cm}
    
    \begin{minipage}{\linewidth}
        \centering
        
        \resizebox{0.65\linewidth}{!}{
            \begin{tabular}{c|cccc}
            \toprule
                Symp. & \multicolumn{4}{c}{Executor}\\
            \cmidrule{2-5}
                & Missing-A & Wrong-A & Inapplicable-A & Wrong-R\\
            \midrule
                Num. & 5 & 4 & \shortstack{\textbf{40}\\(19, 21)} & 4\\
            \midrule
                Perc. & 14.7\% & 11.8\% & \textbf{50.0\%} & 11.8\%\\ 
            \bottomrule
            \end{tabular}
        }
    \end{minipage}
    \vspace{-0.5cm}
    
\end{table}

In the Planner, Missing-P, Malformed-P, Inapplicable-P, and Irrelevant-P each appear in over 14 cases, with Inapplicable-P and Irrelevant-P affecting over 20\% of M-agents.

\rev{The popularity of Inapplicable-P demonstrates the difficulty of understanding the constraints implied in the snapshots.
Among the subcategories, SM has more cases than DM, but we also observe that SM planners are more common than DM planners. LLMs in the DM planner must \textit{identify operable objects}, \textit{locate positions}, \textit{predict action ranges}, and \textit{generate correct tool-calling parameters}. However, the operable objects, positions and action ranges are not explicitly given as prompts, instead, they are implied in the non-textual modality snapshot, bringing challenges to DM planning. LLMs in the SM planner also perform action range prediction and parameter generation despite the summarization of operable objects in the snapshot.}
Among the subcategories of Irrelevant-P, Wrong-Step is more frequent than Wrong-Stop. Its popularity reflects the struggling of M-agents to decompose the general task into multiple unit steps and identify the task end.

\begin{finding}
    \textbf{FINDING 3-2}: \textbf{Inapplicable-P} and \textbf{Irrelevant-P} are the most common symptoms in the \textbf{Planner}.
    Their popularity demonstrate the difficulty in figuring out the \textbf{constraints implied in the non-textual modality snapshots}, decomposing the \textbf{task into small steps} and judging the \textbf{termination} of the task.
\end{finding}


Inapplicable-A dominates the Executor symptoms, accounting for about 76.9\%. 
The popularity of Inapplicable-A indicates that generating effective and correct actions under complex and dynamic multi-modal environments is difficult.
Inapplicable-A is further classified into Wrong-P\&A and Wrong-API. They are almost equally distributed. The popularity of Wrong-P illustrates that many M-agents are designed to follow LLM-generated plans without extra correctness checking. \rev{On the other hand, the commonality of wrong-API symptoms arises from the mismatch between APIs and specific objects. Multi-modal environments are heterogeneous, meaning that objects may have different permissions or interaction interfaces. Wrong-APIs may work on specific, common objects and fail on others.}

\begin{finding}
    \textbf{FINDING 3-3}: \textbf{Inapplicable-A} is the major \textbf{Executor} symptom. 
    Its popularity implies the difficulty to generate \textbf{environment-compatible actions}.
    Its subcategory Wrong-P\&A reflects that M-agents follow LLM-generated plans \textbf{without extra checking}. Wrong-API demonstrates that self-defined APIs may \textbf{fail on special objects}.
\end{finding}

\noindent\textbf{Distribution of root causes in agent-specific implementation bugs.} Table \ref{tab: root cause} displays the distribution of root causes and percentage of M-agents with specific root causes. \rev{Corner-case gaps} and \rev{Error-case gaps} are the major root causes, exhibited in 40 and 38 reports, respectively.

\begin{figure}[!t]
    \centering
    \begin{subfigure}{\linewidth}
        \centering
        \includegraphics[width=\linewidth]{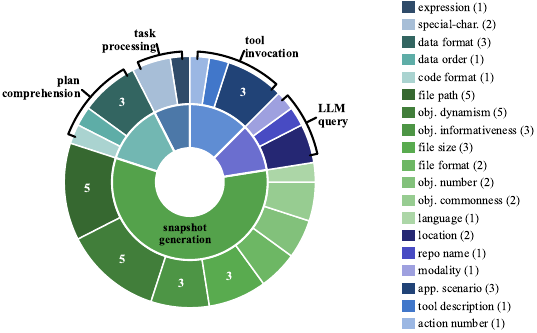}
        \vspace{-0.3cm}
        \caption{Factors related to Corner-cases}
        \label{fig: icch}
    \end{subfigure}
    \begin{subfigure}{\linewidth}
        \centering
        \includegraphics[width=0.9\linewidth]{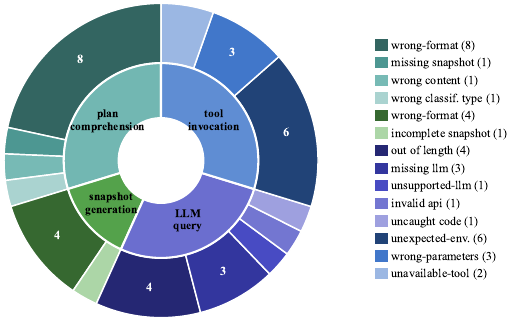}
        \caption{Factors related to Error-cases}
        \vspace{-0.2cm}
        \label{fig: ieh}
    \end{subfigure}
    \caption{\rev{Factors related to Corner-cases and Error-cases distributed by components.}}
    \label{fig: root cause subclasses}
    \vspace{-0.2cm}
\end{figure}

\rev{To figure out the major factors that are overlooked by M-agent developers, Fig. \ref{fig: root cause subclasses} further presents the distribution of factors related to Corner-cases and Error-cases, shown by \ref{fig: icch} and \ref{fig: ieh}, respectively. In Corner-case gaps, snapshot generation dominates the distribution, and its common subclasses include \textit{file path}, \textit{object dynamism}, \textit{file size}, and \textit{object informativeness}. Unlike textual snapshots, non-textual snapshots are mostly saved in files, causing problems when tracking relative file paths or handling LLM-incompatible file sizes. Additionally, while textual snapshots explicitly list operable objects, operable objects in non-textual snapshots are embedded into the background. Perceptors are required but may fail to identify specific objects, such as dynamically generated or running objects, or objects with few functional descriptions.}

\rev{Further more, plan comprehension and tool invocation are the most common components where Error-cases occur, followed by LLM query. Wrong format and unexpected environments are the most frequent factors. Ignorance of wrong format suggests the lacking of format checkers to verify the validness of the LLM generated outputs. The unexpected environments involve various objects with different action ranges, access permissions and subsequent behavior, making it difficult to select appropriate tools and set correct parameters at first try. However, certain tools fail immediately when executed in incompatible environments, yet many M-agents lack explicit mechanisms for handling and analyzing such errors.}



\rev{LLM's limitation} (29.4\%) and \rev{Incorrect semantics} (35.9\%) are also major root causes. \rev{Concurrency, which represents the uncertain sequencing between environment changes and M-agent perceptions or executions, leads to unstable snapshots and inapplicable actions. No persistent memory} causes inconsistent behavior across identical situations.

\begin{table}[!t]
    \centering
    \caption{Distribution of various root causes and percentage of M-agents with specific root causes.}
    \label{tab: root cause}
    \vspace{-0.2cm}
    \resizebox{\linewidth}{!}{
        \begin{tabular}{c|ccccccc}
        \toprule
            Root. & \rev{\shortstack{Concu.}} & \rev{\shortstack{Corner-.}} & \rev{\shortstack{Error-.}} & \rev{\shortstack{No perst.\\mem.}} & \rev{\shortstack{LLM's\\limit.}} & \rev{\shortstack{Bad\\param.}} & \rev{\shortstack{Incor.\\semat.}} \\
        \midrule
            Num. & 1 & \textbf{40} & \textbf{38} & 2 & 33 & 17 & 27\\
        \midrule
            Perc. & 2.9\% & 26.5\% & 17.6\% & 5.9\% & \textbf{29.4\%} & 11.8\% & \textbf{35.9\%}\\
        \bottomrule
        \end{tabular}
    }
    \vspace{-0.2cm}
\end{table}

\begin{finding}
    \textbf{FINDING 4}: \rev{\textbf{Corner-case gaps}} and \rev{\textbf{Error-case gaps}} are the major \textbf{root causes} and cover over 24\% of agent-specific implementation bugs. \rev{The dynamism and diversity of the multi-modal environments bring new factors that are overlooked by M-agent developers, such as the ignorance of \textbf{dynamically generated or running objects} and mishandling of object \textbf{permissions and action ranges}.}
\end{finding}

\subsection{RQ2: Relationship}

\begin{figure}
    \centering
    \includegraphics[width=\linewidth]{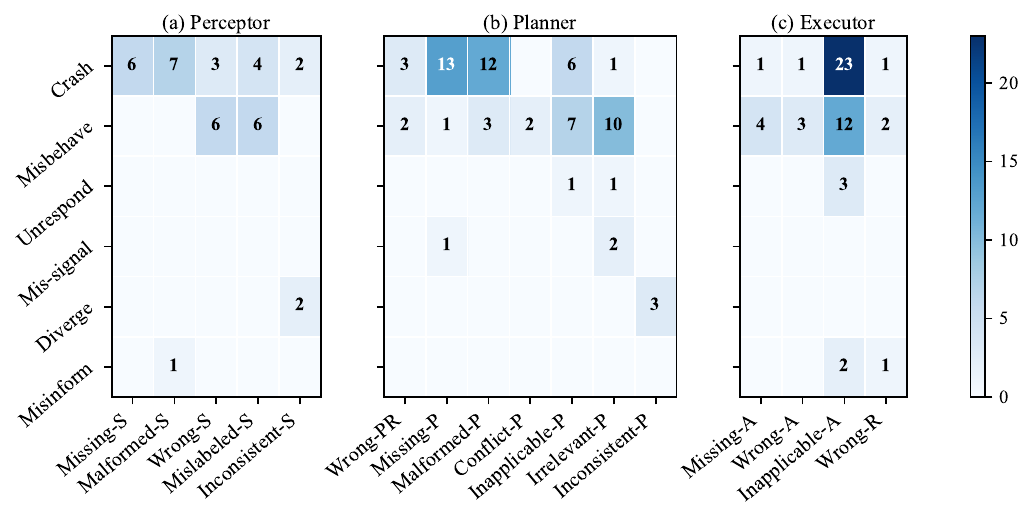}
    \caption{Relationship between the global level symptom and the functionality component level symptoms.}
    \label{fig: func & global}
\end{figure}

\noindent\textbf{Relationship between the global level symptom and the functionality component level symptom.} Fig. \ref{fig: func & global} shows the relationship between the global level symptom and the functionality component related symptoms. Although common global level symptoms like \rev{Crash} and \rev{Misbehave} can be related to multiple functionality component level symptoms, we can still observe patterns from the magnitude of the values. Below, we analyze the results from the perspective of the Perceptor, Planner and Executor, respectively.

In the Perceptor, \rev{Missing-S} and \rev{Malformed-S} are the most common causes of \rev{Crash}, supported by over 6 cases. 
\rev{This connection may be less common in single textual-modal agents, because even if a text snapshot contains missing content or formatting errors, the LLM can usually still parse and process the remaining text. In contrast, multi-modal artifacts such as images or audio files are more susceptible to failures caused by missing file paths or unsuccessful loading, which can prevent the M-agent from accessing the content and even lead to crashes.}

In addition, \rev{Wrong-S} and \rev{Mislabeled-S} generate comprehensible but semantically wrong snapshots, \rev{showing incomplete or outdated environments, or mistakenly marking the uncontrollable objects. Both of them potentially lead to the Planners' misunderstanding of the environment, and further} influence task completion correctness (\rev{Misbehave}). 
\rev{Diverge} is only resulted from \rev{Inconsistent-S} in the Perceptor, because inconsistent snapshots naturally cause inconsistent behavior. The most likely Perceptor related symptom of \rev{Misinform} is \rev{Malformed-S}. \rev{Malformed-S} may be incomprehensible to Planners but comprehensible to the users. Displaying it on the UI without extra feedback misleads users to believe the perception is successful.

\begin{finding}
    \textbf{FINDING 5-1}: \rev{Crash} is likely to be resulted from \textbf{Missing-S} and \textbf{Malformed-S}. \rev{Different from text snapshots, snapshots in image or audio formats are saved in files, making them more susceptible to failures caused by missing file paths or unsuccessful loading, which directly prevent LLMs from accessing the snapshots.} Furthermore, \rev{Misbehave} is connected to \textbf{Wrong-S} and \textbf{Mislabeled-S}, because they generate readable but \textbf{factually incorrect} snapshots, affecting the environmental comprehension.
\end{finding}

In Planner related symptoms, \rev{Missing-P} and \rev{Malformed-P} are most likely to cause \rev{Crash}, supported by around 12 cases. \rev{The Executor is unable to extract valid actions from missing or malformed plans. If the Executor relies on a strict format-matching strategy, it may fail to find the expected content, resulting in errors and ultimately causing a crash.} \rev{Inapplicable-P} and \rev{Irrelevant-P} are highly related to \rev{Misbehave} because these \rev{plans are not executable in the environment or irrelevant to task completion, and often lead to behavioral deviations}. They also cause \rev{Unrespond}: \rev{Executors may be unable to proceed with the subsequent workflow, ultimately resulting in an unresponsive state.} \rev{Mis-signal} is linked to \rev{Irrelevant-P} (the termination signal is part of the plan), and \rev{Diverge} is caused by \rev{Inconsistent-P} as \rev{local inconsistencies may lead to overall inconsistencies in execution results.}

\begin{finding}
    \textbf{FINDING 5-2}: In Planner related symptoms, \rev{Crash} is related to \rev{Missing-P} and \rev{Malformed-P because a part of Executors adopt template-matching strategies, and they may raise errors when encountering plans that do not match the expected format.} \rev{Misbehave} is connected to \rev{Inapplicable-P} and \rev{Irrelevant-P}. We also observe that \rev{Unrespond} and \rev{Mis-signal} are resulted from Planner symptoms.
\end{finding}


In Executor related symptoms, \rev{Crash}, \rev{Misbehave} and \rev{Unrespond} are highly connected to \rev{Inapplicable-A}, supported by 23, 12 and 3 cases, respectively. \rev{The type of global symptoms depends on the different manifestations of Inapplicable-A.} \rev{In text-based environments, actions are typically applied to a well-defined object (e.g., code). In contrast, in multi-modal environments, the objects are more diverse and dynamic, so an action may attempt to access a non-existent object. In such cases, Crash} most likely happens. \rev{In other cases,} the \rev{Inapplicable-A} may call wrong implemented API and \rev{Misbehave} is more prone to occur. \rev{Misinform} is related to \rev{Inapplicable-A} and \rev{Wrong-R because users may be incorrectly informed that the \rev{Inapplicable-A} is successfully executed}.

\begin{finding}
    \textbf{FINDING 5-3}: \rev{Crash}, \rev{Misbehave} and \rev{Unrespond} are all connected to \textbf{Inapplicable-A} in the Executor. The implementation of the actions determines which global level symptoms occur. For example, actions can access \textbf{unavailable objects} \rev{in the dynamic and diverse multi-modal environments and cause Crash}. Otherwise, if \rev{Inapplicable-A} calls \textbf{wrong-implemented APIs}, \rev{Misbehave} happens.
\end{finding}

\begin{figure}
    \centering
    \includegraphics[width=\linewidth]{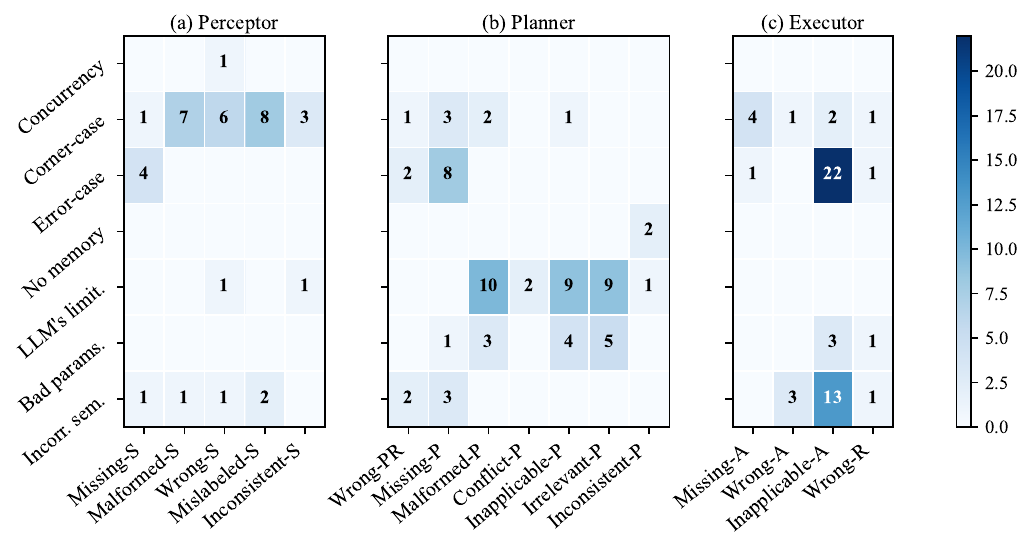}
    \caption{Relationship between the functionality component related symptoms and the root causes.}
    \label{fig: func & root}
\end{figure}

\textbf{Relationship between the functionality component level symptom and the root cause.} Fig. \ref{fig: func & root} illustrates the relationship between root causes and functionality component level symptoms. 
In the Perceptor, \rev{Malformed-S}, \rev{Wrong-S} and \rev{Mislabeled-S} are all related to \rev{Corner-case gaps}, each relationship supported by over 3 cases. 
\rev{After further analyzing the subcategories of Corner-case, we find that all 23 Perceptor symptoms caused by Corner-case gaps can be attributed to the snapshot generation component within Perceptor. Only 2 Malformed-S cases need to be traced back to the earlier task processing stage. In these 2 exceptional cases, we find that the task name is directly used as part of the snapshot filename without any sanitization. As a result, the filename contains special characters from the task name that are not permitted in filenames.} Additionally, \rev{Wrong-S} and \rev{Mislabeled-S} are mostly resulted from bad handling of \rev{object dynamism} and \rev{object informativeness}.

\begin{finding}
    \textbf{FINDING 6-1}: \rev{Malformed-S}, \rev{Wrong-S} and \rev{Mislabeled-S} are connected to \textbf{Corner-case gaps} because of \rev{mishandling \textbf{object informativeness} or \textbf{object dynamism} in the snapshot generation or ignoring \textbf{special characters} when using task content as the snapshot filename.}
\end{finding}

In Planner symptoms, \rev{Malformed-P}, \rev{Conflict-P}, \rev{Inapplicable-P} and \rev{Irrelevant-P} are all connected to \rev{LLM's Limitation}. Due to \rev{internal limitations} like hallucinations, LLM-generated plans may violate output format, environmental constraints, or task descriptions. \rev{Wrong-PR} traces to \rev{Error-case gaps} (\rev{unhandled prompt length overflow in the LLM query component}), and \rev{Missing-P} also stems from \rev{Error-case gaps} (e.g., \rev{invalid API, unsupported LLM, missing LLM}). \rev{Inconsistent-P} is connected to \rev{No persistent memory}, as memory records history needed to maintain behavioral consistency.

\begin{finding}
    \textbf{FINDING 6-2}: \rev{Malformed-P}, \rev{Conflict-P}, \rev{Inapplicable-P} and \rev{irrelevant-P} are connected to \textbf{LLM's Limitation} because LLM-generated plan is \textbf{not trustworthy}, no matter in output format, internal logic, environment applicability or task relation. \rev{Wrong-PR} is highly related to \rev{Error-case gaps} as wrong prompts can be caused by \textbf{unhandled length excess}. \rev{Inconsistent-P} is connected to \rev{No persistent memory} as memory maintains behavior consistency.
\end{finding}



In Executor symptoms, \rev{Missing-A} is related to \rev{Corner-case gaps}. \rev{Limited support for diverse data formats and tools with incomplete descriptions cause inappropriate generation of actions.} \rev{Wrong-A} is caused by \rev{Incorrect semantics} due to incorrect plan-to-action parsing. \rev{Inapplicable-A} is caused by both \rev{Error-case gaps} (unchecked access to unavailable objects in \rev{unexpected environments}) and \rev{Incorrect semantics} (incorrect API implementations).

\begin{finding}
    \textbf{FINDING 6-3}: \rev{Missing-A} is related to \rev{Corner-case gaps} due to \rev{mishandling of plans with diverse formats and tools with incomplete descriptions}. \rev{Wrong-A} is likely to be caused by \rev{Incorrect semantics} owing to incorrect parsing from plans to actions. Finally, \rev{Inapplicable-A} is highly connected to \textbf{Error-case gaps} and \textbf{Incorrect semantics}, manifesting as \textbf{ignoring unavailable object access}, and \textbf{inconsistent API implementation and description}.
\end{finding}


\subsection{RQ3: Applicability}

\subsubsection{Implementation of \sysname}
\input{new/tester}

%
%
%
%
\subsubsection{Settings of \sysname}

\mbox{}\\
\noindent \textbf{Implementation.} The judging LLM is GPT-4o, a multi-modal LLM, enabling inter-modality and cross-modality consistency checks. Each task is queried 3 times and the majority verdict is used. Temperature is set to 0.2 for stable results.

\noindent \textbf{Benchmarks.} The benchmark consists of seven randomly selected M-agents from GitHub, manually instrumented for reproducibility.

\noindent \textbf{Testcases.} Test cases specify goal-oriented tasks tailored to each agent's functionality, using original tasks augmented with ChatGPT-generated mutations. The test suite comprises 65 tasks spanning four application scenarios.

\noindent \textbf{Constraints.} The M-agents assigned with a specific task should finish the task in 5 minutes and 20 rounds. \rev{These constraints are set with reference to human completion time. Humans require about 10 steps to complete each of these tasks. Agents that cannot conduct a task in 20 steps are considered as incompetent. Normally, 20 consecutive rounds are finished in 3 minutes. Agents that cannot conduct 20 rounds in 5 minutes are mostly unresponsive.} In addition, each task is run at least twice. For M-agents without additional assigned tasks, such as autonomous driving M-agents, they are forced to stop after running for 5 minutes. These M-agents run for at least 20 times and each time the environment is randomly initialized.

The applicability evaluation was run on the Ubuntu 20.04.3 machine with Intel$^\circledR$ Core$^\text{TM}$ i7-10700 Processor CPU@2.9GHz and 16GB RAM. 
The \textit{source code}, \textit{datasets} and \textit{case studies} are available at \textbf{our website} \cite{tool}.

\subsubsection{\sysname's Discovery}

\sysname discovers 266 bugs in total. After deduplication, \rev{41} unique bugs remain. \rev{Manual validation following the procedure in Section \ref{sec: labeling} finds one mis-classified symptom, giving a 97.6\% accuracy rate. The 40 confirmed unique bugs cover 61.4\% of open issues and include 31 new bugs}, demonstrating the effectiveness of \sysname.
Table \ref{tab: symptom_7_agent} shows the symptom distribution of unique bugs. Crash and Misbehave dominate. All components are covered with at least 19 reports. Wrong-S, Irrelevant-P, and Inapplicable-A are the leading symptoms in the Perceptor, Planner, and Executor respectively, consistent with our study.

\begin{table}[!t]
    \centering
    \caption{\sysname's discovery: covered open issues and new bugs. The row ``coverage'' records the percentage of open issues that are covered by \sysname. The value ``-'' represents no open issues.}
    \label{tab: output}
    \vspace{-0.2cm}
    \resizebox{\linewidth}{!}{
        \begin{tabular}{c|cccccccccccc|c}
        \toprule
            id & \#1 & \#2 & \#3 & \#4 & \#5 & \#6 & \#7 & \#8 & \#9 & \#10 & \#11 & \#12 & summary\\
        \midrule
            total & 20 & 45 & 61 & 53 & 18 & 20 & 9 & 16 & 3 & 7 & 7 & 7 & 266 (sum.)\\
        \midrule
            unique & 6 & 4 & 7 & 4 & 5 & 1 & 3 & 2 & 1 & 3 & 2 & 2 & 40 (sum.)\\
        \midrule
            new & 6 & 1 & 3 & 4 & 5 & 1 & 3 & 2 & 1 & 3 & 1 & 1 & 31 (sum.)\\
        \midrule
            cover. & - & 66.7\% & 61.5\% & - & - & - & - & - & - & - & 50\% & 57.1\% & 61.4\% (ave.)\\
        \bottomrule
        \end{tabular}
    }
    \vspace{-0.2cm}
\end{table}


\begin{table}[!t]
    \centering
    \caption{Distribution of symptoms identified by \sysname on 12 agents.}
    \label{tab: symptom_7_agent}
    \vspace{-0.2cm}
    \resizebox{0.9\linewidth}{!}{
    \begin{tabular}{cccccc|ccc}
    \toprule
        \multicolumn{6}{c|}{Glob.} & \multicolumn{3}{c}{Func.}\\
    \midrule
        Crash & Misb. & Unresp. & Mis-sig. & Diverge & Misinf. & Percept. & Plann. & Execut. \\
    \midrule
        14 & 15 & 3 & 6 & 2 & 0 & 27 & 29 & 14\\
    \bottomrule
    \end{tabular}
    }
\vspace{-0.2cm}
\end{table}

\begin{finding}
    \textbf{FINDING 7}: \sysname's discovery covers \textbf{61.4\% of open issues} and includes \textbf{31 new bugs}. The findings are consistent with our study. Wrong-S, Irrelevant-P, Inapplicable-A are the most common symptoms in the three components, respectively.
\end{finding}

%% file: new/tester.tex
We demonstrate the usefulness of our finding by implementing a preliminary tester called \sysname based on our taxonomy in Section \ref{sec: taxonomy}. \sysname can automatically identify global level symptom and localize the functionality component level symptom based on the M-agent's inter-component outputs.


We manually \textit{instrumented} each M-agent to emit formatted inter-component outputs (environments, snapshots, prompts, plans, actions, and reflects) each round, plus a summary file on termination. Test datasets were constructed by \textit{mutating original tasks} and randomly varying initial environments to assess Planner generality and Executor adaptability.

\begin{figure}[!t]
    \centering
    \includegraphics[width=0.98\linewidth]{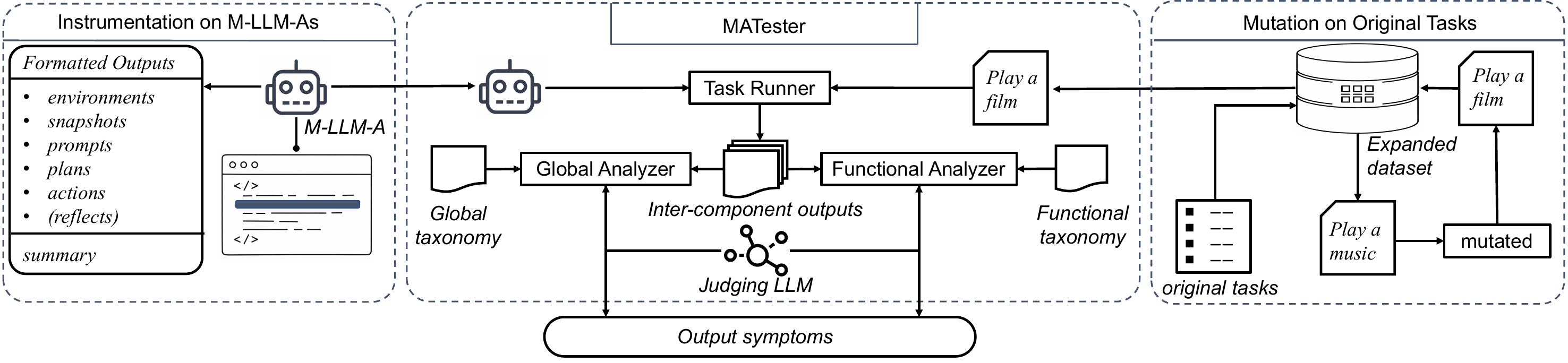}
    \vspace{-0.2cm}
    \caption{The workflow of \sysname.}
    \label{fig: workflow}
    \vspace{-0.5cm}
\end{figure}


\sysname tests an M-agent in three steps as shown in Fig. \ref{fig: workflow}. First, the \textit{Task Runner} invokes the agent and enforces time and round limits. Then the \textit{Global Analyzer} identifies global symptoms: Crash and Unrespond are detected from error reports and hanging time; Diverge is detected by running similar tasks twice and comparing outcomes; Misbehave and Mis-signal are judged by a LLM that checks task completeness and misbehavior against the environment and task description.


If a global symptom is found, the \textit{Functional Analyzer} localizes component-level symptoms (Section \ref{sec: functionality symptom}). For each component, it first checks output existence and format (detecting Missing-S, Malformed-S, etc.), then uses the judging LLM to compare snapshot/plan/action contents against the environment and task, detecting semantic symptoms such as Missing-S, Mislabeled-S, and Inconsistent-S. The core of this comparison is identifying missing, wrong, or conflicted information across multiple modalities, which is suited to the M-LLM's cross-modal reasoning capability.

%% file: new/discussion.tex
\section{Discussion}

\subsection{Existing Adopted Strategies to Mitigate M-agent Bugs}

We observe that developers adopt ad-hoc mitigation strategies that are often insufficient. For instance, using LLMs to handle dynamic objects still produces Wrong-S and Inconsistent-S due to LLMs' instability, causing Misbehave and Diverge. Similarly, setting a default fallback action to avoid Missing-A can perpetuate mistakes and lead to Misbehave. These strategies are too simple to resolve the underlying bugs, and can introduce side effects. Our study reveals that symptoms stem from multiple interacting root causes, requiring more rigorous and agent-specific designs.

\subsection{Suggestions to Developers and Users}

\noindent\textbf{To developers.}
For the \textit{Perceptor}, \textit{support diverse objects, especially dynamic objects and less informative objects}. Combine static layout analysis with visual observer to capture both static and dynamic objects. For the \textit{Planner}, \textit{remain skeptical of LLM outputs}. Validate format, environmental constraints, task relevance, and use self-reflection and memory to improve consistency. For the \textit{Executor}, \textit{establish a complete error-handling workflow}. Verify action format and object availability before execution, and feed errors back to the Planner. Ensure API semantics match their names and descriptions.

\noindent\textbf{To users.}
M-agents are unreliable in safety-critical scenarios and may report actions inconsistently with what they actually perform. Users should exercise caution to avoid financial or privacy leakage, and verify agent behavior directly rather than relying on its self-reported feedback.

\subsection{Threats to Validity}
Three threats may affect validity. First, our manual classification and labeling process may introduce subjectivity despite inter-annotator agreement checks. Second, the selected M-agents and issues may not cover all application scenarios. New design paradigms may yield undiscovered symptoms and root causes. Third, \sysname analyzes inter-component outputs and may miss intra-component bugs. Future work will apply static source-code analysis to address this gap.

%% file: new/related-work.tex
\section{Related Work}
\noindent\textbf{Multi-modal LLM-based Agents.} 
M-agents have emerged as a promising research topic. 
One of the most popular directions is GUI automation.
Typical research include WebWISE \cite{WebWISE}, AutoDroid \cite{AutoDroid}, MM-Navigator \cite{MM-Navigator}, Auto-GUI \cite{Auto-GUI}, Appagent \cite{AppAgent}, DroidBot-GPT \cite{DroidBot-GPT}, MobileGPT \cite{MobileGPT}, Assistgui \cite{ASSISTGUI} and WebExperT \cite{WebExperT}. M-agents are also developed in open-world environments, such as MP5 \cite{MP5}, JARVIS-1 \cite{JARVIS-1}, Emma-Alfworld \cite{Emma-Alfworld}, MC-Planner \cite{MC-Planner}, Octopus \cite{Octopus}, STEVE \cite{STEVE} and LLM-REGRESS \cite{LLM-REGRESS}. Other widely studied areas include autonomous driving \cite{GPT-Driver,DriveLikeAHuman,GPT4V-AD-Exploration}, navigation \cite{CityNavAgent,RILA}, visual generation \cite{LLaVA-Interactive,Visual-ChatGPT,GenArtist} and audio generation \cite{Loop-Copilot,MusicAgent,WavJourney}.

\noindent\textbf{Bug Study on LLM-based Agents.} 
The development of agents brings new concerns. Several empirical studies have been conducted to summary and analyze their security \cite{empirical_security}, compliance \cite{empirical_compliance}, \rev{code-level implementation defects \cite{agent_fix_agent} and general module-level bugs \cite{llm_agent_module}}. Specifically, Rahardja et al. \cite{agent_fix_agent} analyzed 201 agent bugs and introduced a taxonomy. This taxonomy is at the code level, related to dependency, configuration and initialization. \rev{Cemri et al. \cite{llm_agent_module} summarized bugs at the general module level, related to memory, reflection, planning, action and system}. There are also studies of agents \rev{in specific application domains}, such as Software Engineering Agents \cite{understand_SWEA,understand_SWEA_2}, \rev{Code Generation Agents \cite{understand_codeA}} and Search Agents \cite{understand_SA}. Compared with them, \rev{we focused on M-agents. We classified and analyzed common bugs happened when interacting with the real-time and heterogeneous multi-modal environments. We proposed three levels of symptoms and root causes, and analyzed commonly ignored factors.}

%% file: new/conclusion.tex
\section{Conclusion}
M-agents have been widely applied to real-world safety-critical scenarios. Their interaction environments are mostly real-time and heterogeneous, bringing challenges to the design of M-agents.
In this paper, we proposed the first study on the bugs in M-agents. Specifically, we analyzed 34 M-agents and summarized a top-down taxonomy to include end-user-observable global symptoms, developer-oriented component symptoms and root causes. Guided by the taxonomy, we implemented \sysname to automatically identify bugs. When applying to 12 extra M-agents, \sysname covers 61.4\% of open issues and discovers 31 new bugs, which proves the usefulness of our findings. This paper provides a guidance for further development and usage of M-agents.

%% file: reference.bib
@article{consel_agent,
  author       = {Huachuan Qiu and
                  Zhenzhong Lan},
  title        = {Interactive Agents: Simulating Counselor-Client Psychological Counseling
                  via Role-Playing LLM-to-LLM Interactions},
  journal      = {CoRR},
  volume       = {abs/2408.15787},
  year         = {2024},
  doi          = {10.48550/ARXIV.2408.15787},
  eprinttype    = {arXiv},
  eprint       = {2408.15787},
  bibsource    = {dblp computer science bibliography, https://dblp.org}
}

@misc{traffic_accident,
  author       = {U.S National Transportation Safety Board},
  title        = {Collision Between Vehicle Controlled by Developmental Automated Driving System and Pedestrian},
  howpublished = {\url{www.ntsb.gov/investigations/
AccidentReports/Reports/HAR1903.pdf}},
  year         = {2019}
}

@misc{financial_loss,
  author       = {Blockchain News},
  title        = {Microsoft's AI agent spends 100\% of its testing funds on online fraud—Lessons for MSFT and AI-secured transactions.},
  howpublished = {\url{https://blockchain.news/zh/flashnews/microsoft-ai-agents-spent-100-of-test-funds-on-online-scams-trading-takeaways-for-msft-and-ai-security-plays-zh}},
  year         = {2025}
}

@misc{privacy_leakage,
  author       = {Netmag},
  title        = {ChatGPT agent exposes vulnerability, potentially allowing ``Shadow Leak''' to steal sensitive Gmail data.},
  howpublished = {\url{https://netmag.tw/2025/09/24/chatgpt-deep-research-gmail-security}},
  year         = {2025}
}

@misc{tool,
  author       = {MATester},
  title        = {MATester.},
  howpublished = {\url{https://github.com/MATester0/MATester}},
  year         = {2026}
}

@article{SWE_agent_1,
  author       = {Chunqiu Steven Xia and
                  Yinlin Deng and
                  Soren Dunn and
                  Lingming Zhang},
  title        = {Agentless: Demystifying LLM-based Software Engineering Agents},
  journal      = {CoRR},
  volume       = {abs/2407.01489},
  year         = {2024},
  doi          = {10.48550/ARXIV.2407.01489},
  eprinttype    = {arXiv},
  eprint       = {2407.01489},
  bibsource    = {dblp computer science bibliography, https://dblp.org}
}

@inproceedings{SWE_agent_2,
  author       = {Yuntong Zhang and
                  Haifeng Ruan and
                  Zhiyu Fan and
                  Abhik Roychoudhury},
  editor       = {Maria Christakis and
                  Michael Pradel},
  title        = {AutoCodeRover: Autonomous Program Improvement},
  booktitle    = {Proceedings of the 33rd {ACM} {SIGSOFT} International Symposium on
                  Software Testing and Analysis, {ISSTA} 2024, Vienna, Austria, September
                  16-20, 2024},
  pages        = {1592--1604},
  publisher    = {{ACM}},
  year         = {2024},
  doi          = {10.1145/3650212.3680384},
  bibsource    = {dblp computer science bibliography, https://dblp.org}
}

@inproceedings{SWE_agent_3,
  author       = {Xingyao Wang and
                  Boxuan Li and
                  Yufan Song and
                  Frank F. Xu and
                  Xiangru Tang and
                  Mingchen Zhuge and
                  Jiayi Pan and
                  Yueqi Song and
                  Bowen Li and
                  Jaskirat Singh and
                  Hoang H. Tran and
                  Fuqiang Li and
                  Ren Ma and
                  Mingzhang Zheng and
                  Bill Qian and
                  Yanjun Shao and
                  Niklas Muennighoff and
                  Yizhe Zhang and
                  Binyuan Hui and
                  Junyang Lin and
                  et al.},
  title        = {OpenHands: An Open Platform for {AI} Software Developers as Generalist
                  Agents},
  booktitle    = {The Thirteenth International Conference on Learning Representations,
                  {ICLR} 2025, Singapore, April 24-28, 2025},
  publisher    = {OpenReview.net},
  year         = {2025},
  bibsource    = {dblp computer science bibliography, https://dblp.org}
}

@article{reason_agent_1,
  author       = {Xiao Wang and
                  Jia Wang and
                  Yijie Wang and
                  Pengtao Dang and
                  Sha Cao and
                  Chi Zhang},
  title        = {{MARS:} toward more efficient multi-agent collaboration for {LLM}
                  reasoning},
  journal      = {CoRR},
  volume       = {abs/2509.20502},
  year         = {2025},
  doi          = {10.48550/ARXIV.2509.20502},
  eprinttype    = {arXiv},
  eprint       = {2509.20502},
  bibsource    = {dblp computer science bibliography, https://dblp.org}
}

@inproceedings{reason_agent_2,
  author       = {Junde Wu and
                  Jiayuan Zhu and
                  Yuyuan Liu and
                  Min Xu and
                  Yueming Jin},
  editor       = {Wanxiang Che and
                  Joyce Nabende and
                  Ekaterina Shutova and
                  Mohammad Taher Pilehvar},
  title        = {Agentic Reasoning: {A} Streamlined Framework for Enhancing {LLM} Reasoning
                  with Agentic Tools},
  booktitle    = {Proceedings of the 63rd Annual Meeting of the Association for Computational
                  Linguistics (Volume 1: Long Papers), {ACL} 2025, Vienna, Austria,
                  July 27 - August 1, 2025},
  pages        = {28489--28503},
  publisher    = {Association for Computational Linguistics},
  year         = {2025},
  bibsource    = {dblp computer science bibliography, https://dblp.org}
}

@article{large_language_model_verilog,
 author={Guang Yang and Wei Zheng and Xiang Chen and Dong Liang and Peng Hu and Yukui Yang and Shaohang Peng and Zhenghan Li and Jiahui Feng and Xiao Wei and Kexin Sun and Deyuan Ma and Haotian Cheng and Yiheng Shen and Xing Hu and Terry Yue Zhuo and David Lo},
 title={Large Language Model for Verilog Code Generation: Literature Review and the Road Ahead}, 
 journal      = {CoRR},
 volume       = {abs/2512.00020},
 year={2025},
 doi          = {10.48550/ARXIV.2512.00020},
 eprinttype    = {arXiv},
 eprint={2512.00020},
 bibsource    = {dblp computer science bibliography, https://dblp.org}
}

@article{M-LLM-A_survey,
  author       = {Junlin Xie and
                  Zhihong Chen and
                  Ruifei Zhang and
                  Xiang Wan and
                  Guanbin Li},
  title        = {Large Multimodal Agents: {A} Survey},
  journal      = {CoRR},
  volume       = {abs/2402.15116},
  year         = {2024},
  doi          = {10.48550/ARXIV.2402.15116},
  eprinttype    = {arXiv},
  eprint       = {2402.15116},
  bibsource    = {dblp computer science bibliography, https://dblp.org}
}

@inproceedings{vehicle_bugs_study,
  author       = {Joshua Garcia and
                  Yang Feng and
                  Junjie Shen and
                  Sumaya Almanee and
                  Yuan Xia and
                  Qi Alfred Chen},
  editor       = {Gregg Rothermel and
                  Doo{-}Hwan Bae},
  title        = {A comprehensive study of autonomous vehicle bugs},
  booktitle    = {{ICSE} '20: 42nd International Conference on Software Engineering,
                  Seoul, South Korea, 27 June - 19 July, 2020},
  pages        = {385--396},
  publisher    = {{ACM}},
  year         = {2020},
  doi          = {10.1145/3377811.3380397},
  bibsource    = {dblp computer science bibliography, https://dblp.org}
}

@article{deep_learning_program_study,
  author       = {Amin Nikanjam and
                  Mohammad Mehdi Morovati and
                  Foutse Khomh and
                  Houssem Ben Braiek},
  title        = {Faults in deep reinforcement learning programs: a taxonomy and a detection
                  approach},
  journal      = {Autom. Softw. Eng.},
  volume       = {29},
  number       = {1},
  pages        = {8},
  year         = {2022},
  doi          = {10.1007/S10515-021-00313-X},
  bibsource    = {dblp computer science bibliography, https://dblp.org}
}

@inproceedings{deep_learning_compiler_study,
  author       = {Qingchao Shen and
                  Haoyang Ma and
                  Junjie Chen and
                  Yongqiang Tian and
                  Shing{-}Chi Cheung and
                  Xiang Chen},
  editor       = {Diomidis Spinellis and
                  Georgios Gousios and
                  Marsha Chechik and
                  Massimiliano Di Penta},
  title        = {A comprehensive study of deep learning compiler bugs},
  booktitle    = {{ESEC/FSE} '21: 29th {ACM} Joint European Software Engineering Conference
                  and Symposium on the Foundations of Software Engineering, Athens,
                  Greece, August 23-28, 2021},
  pages        = {968--980},
  publisher    = {{ACM}},
  year         = {2021},
  doi          = {10.1145/3468264.3468591},
  bibsource    = {dblp computer science bibliography, https://dblp.org}
}

@article{QGS,
  author       = {He Zhang and
                  Muhammad Ali Babar and
                  Paolo Tell},
  title        = {Identifying relevant studies in software engineering},
  journal      = {Inf. Softw. Technol.},
  volume       = {53},
  number       = {6},
  pages        = {625--637},
  year         = {2011},
  doi          = {10.1016/J.INFSOF.2010.12.010},
  bibsource    = {dblp computer science bibliography, https://dblp.org}
}

@inproceedings{snowballing,
  author       = {Claes Wohlin},
  editor       = {Martin J. Shepperd and
                  Tracy Hall and
                  Ingunn Myrtveit},
  title        = {Guidelines for snowballing in systematic literature studies and a
                  replication in software engineering},
  booktitle    = {18th International Conference on Evaluation and Assessment in Software
                  Engineering, {EASE} '14, London, England, United Kingdom, May 13-14,
                  2014},
  pages        = {38:1--38:10},
  publisher    = {{ACM}},
  year         = {2014},
  doi          = {10.1145/2601248.2601268},
  bibsource    = {dblp computer science bibliography, https://dblp.org}
}

@inproceedings{issue_classifier,
  author       = {Maliheh Izadi},
  editor       = {Andrea Di Sorbo and
                  Sebastiano Panichella},
  title        = {CatIss: An Intelligent Tool for Categorizing Issues Reports using
                  Transformers},
  booktitle    = {2022 {IEEE/ACM} 1st International Workshop on Natural Language-Based
                  Software Engineering {(NLBSE} 2022), Co-located with {ICSE} 2022,
                  Pittsburgh, PA, USA, May 8, 2022},
  pages        = {44--47},
  publisher    = {{ACM/IEEE}},
  year         = {2022},
  doi          = {10.1145/3528588.3528662},
  bibsource    = {dblp computer science bibliography, https://dblp.org}
}

@inproceedings{issue_discussion_classifier,
  author       = {Deeksha M. Arya and
                  Wenting Wang and
                  Jin L. C. Guo and
                  Jinghui Cheng},
  editor       = {Joanne M. Atlee and
                  Tevfik Bultan and
                  Jon Whittle},
  title        = {Analysis and detection of information types of open source software
                  issue discussions},
  booktitle    = {Proceedings of the 41st International Conference on Software Engineering,
                  {ICSE} 2019, Montreal, QC, Canada, May 25-31, 2019},
  pages        = {454--464},
  publisher    = {{IEEE} / {ACM}},
  year         = {2019},
  doi          = {10.1109/ICSE.2019.00058},
  bibsource    = {dblp computer science bibliography, https://dblp.org}
}

@article{bug_characteristics,
  author       = {Lin Tan and
                  Chen Liu and
                  Zhenmin Li and
                  Xuanhui Wang and
                  Yuanyuan Zhou and
                  ChengXiang Zhai},
  title        = {Bug characteristics in open source software},
  journal      = {Empir. Softw. Eng.},
  volume       = {19},
  number       = {6},
  pages        = {1665--1705},
  year         = {2014},
  doi          = {10.1007/S10664-013-9258-8},
  bibsource    = {dblp computer science bibliography, https://dblp.org}
}

@inproceedings{Cohen_s_kappa_coefficient,
  author       = {Susana M. Vieira and
                  Uzay Kaymak and
                  Jo{\~{a}}o M. C. Sousa},
  title        = {Cohen's kappa coefficient as a performance measure for feature selection},
  booktitle    = {{FUZZ-IEEE} 2010, {IEEE} International Conference on Fuzzy Systems,
                  Barcelona, Spain, 18-23 July, 2010, Proceedings},
  pages        = {1--8},
  publisher    = {{IEEE}},
  year         = {2010},
  doi          = {10.1109/FUZZY.2010.5584447},
  bibsource    = {dblp computer science bibliography, https://dblp.org}
}

@inproceedings{WebWISE,
  author       = {Heyi Tao and
                  Sethuraman TV and
                  Michal Shlapentokh{-}Rothman and
                  Tanmay Gupta and
                  Heng Ji and
                  Derek Hoiem},
  editor       = {Kevin Duh and
                  Helena G{\'{o}}mez{-}Adorno and
                  Steven Bethard},
  title        = {WebWISE: Unlocking Web Interface Control for LLMs via Sequential Exploration},
  booktitle    = {Findings of the Association for Computational Linguistics: {NAACL}
                  2024, Mexico City, Mexico, June 16-21, 2024},
  pages        = {3693--3711},
  publisher    = {Association for Computational Linguistics},
  year         = {2024},
  doi          = {10.18653/V1/2024.FINDINGS-NAACL.234},
  bibsource    = {dblp computer science bibliography, https://dblp.org}
}

@article{AutoDroid,
  author       = {Hao Wen and
                  Yuanchun Li and
                  Guohong Liu and
                  Shanhui Zhao and
                  Tao Yu and
                  Toby Jia{-}Jun Li and
                  Shiqi Jiang and
                  Yunhao Liu and
                  Yaqin Zhang and
                  Yunxin Liu},
  title        = {Empowering {LLM} to use Smartphone for Intelligent Task Automation},
  journal      = {CoRR},
  volume       = {abs/2308.15272},
  year         = {2023},
  doi          = {10.48550/ARXIV.2308.15272},
  eprinttype    = {arXiv},
  eprint       = {2308.15272},
  bibsource    = {dblp computer science bibliography, https://dblp.org}
}

@article{MM-Navigator,
  author       = {An Yan and
                  Zhengyuan Yang and
                  Wanrong Zhu and
                  Kevin Lin and
                  Linjie Li and
                  Jianfeng Wang and
                  Jianwei Yang and
                  Yiwu Zhong and
                  Julian J. McAuley and
                  Jianfeng Gao and
                  Zicheng Liu and
                  Lijuan Wang},
  title        = {{GPT-4V} in Wonderland: Large Multimodal Models for Zero-Shot Smartphone
                  {GUI} Navigation},
  journal      = {CoRR},
  volume       = {abs/2311.07562},
  year         = {2023},
  doi          = {10.48550/ARXIV.2311.07562},
  eprinttype    = {arXiv},
  eprint       = {2311.07562},
  bibsource    = {dblp computer science bibliography, https://dblp.org}
}

@inproceedings{Auto-GUI,
  author       = {Zhuosheng Zhang and
                  Aston Zhang},
  editor       = {Lun{-}Wei Ku and
                  Andre Martins and
                  Vivek Srikumar},
  title        = {You Only Look at Screens: Multimodal Chain-of-Action Agents},
  booktitle    = {Findings of the Association for Computational Linguistics, {ACL} 2024,
                  Bangkok, Thailand and virtual meeting, August 11-16, 2024},
  pages        = {3132--3149},
  publisher    = {Association for Computational Linguistics},
  year         = {2024},
  doi          = {10.18653/V1/2024.FINDINGS-ACL.186},
  bibsource    = {dblp computer science bibliography, https://dblp.org}
}

@inproceedings{AppAgent,
  author       = {Chi Zhang and
                  Zhao Yang and
                  Jiaxuan Liu and
                  Yanda Li and
                  Yucheng Han and
                  Xin Chen and
                  Zebiao Huang and
                  Bin Fu and
                  Gang Yu},
  editor       = {Naomi Yamashita and
                  Vanessa Evers and
                  Koji Yatani and
                  Sharon Xianghua Ding and
                  Bongshin Lee and
                  Marshini Chetty and
                  Phoebe O. Toups Dugas},
  title        = {AppAgent: Multimodal Agents as Smartphone Users},
  booktitle    = {Proceedings of the 2025 {CHI} Conference on Human Factors in Computing
                  Systems, {CHI} 2025, YokohamaJapan, 26 April 2025- 1 May 2025},
  pages        = {70:1--70:20},
  publisher    = {{ACM}},
  year         = {2025},
  doi          = {10.1145/3706598.3713600},
  bibsource    = {dblp computer science bibliography, https://dblp.org}
}

@article{DroidBot-GPT,
  author       = {Hao Wen and
                  Hongming Wang and
                  Jiaxuan Liu and
                  Yuanchun Li},
  title        = {DroidBot-GPT: GPT-powered {UI} Automation for Android},
  journal      = {CoRR},
  volume       = {abs/2304.07061},
  year         = {2023},
  doi          = {10.48550/ARXIV.2304.07061},
  eprinttype    = {arXiv},
  eprint       = {2304.07061},
  bibsource    = {dblp computer science bibliography, https://dblp.org}
}

@inproceedings{MobileGPT,
  author       = {Sunjae Lee and
                  Junyoung Choi and
                  Jungjae Lee and
                  Munim Hasan Wasi and
                  Hojun Choi and
                  Steven Y. Ko and
                  Sangeun Oh and
                  Insik Shin},
  editor       = {Weisong Shi and
                  Deepak Ganesan and
                  Nicholas D. Lane},
  title        = {MobileGPT: Augmenting {LLM} with Human-like App Memory for Mobile
                  Task Automation},
  booktitle    = {Proceedings of the 30th Annual International Conference on Mobile
                  Computing and Networking, {ACM} MobiCom 2024, Washington D.C., DC,
                  USA, November 18-22, 2024},
  pages        = {1119--1133},
  publisher    = {{ACM}},
  year         = {2024},
  doi          = {10.1145/3636534.3690682},
  bibsource    = {dblp computer science bibliography, https://dblp.org}
}

@article{ASSISTGUI,
  author       = {Difei Gao and
                  Lei Ji and
                  Zechen Bai and
                  Mingyu Ouyang and
                  Peiran Li and
                  Dongxing Mao and
                  Qinchen Wu and
                  Weichen Zhang and
                  Peiyi Wang and
                  Xiangwu Guo and
                  Hengxu Wang and
                  Luowei Zhou and
                  Mike Zheng Shou},
  title        = {{ASSISTGUI:} Task-Oriented Desktop Graphical User Interface Automation},
  journal      = {CoRR},
  volume       = {abs/2312.13108},
  year         = {2023},
  doi          = {10.48550/ARXIV.2312.13108},
  eprinttype    = {arXiv},
  eprint       = {2312.13108},
  bibsource    = {dblp computer science bibliography, https://dblp.org}
}

@inproceedings{WebExperT,
  author       = {Haohao Luo and
                  Jiayi Kuang and
                  Wei Liu and
                  Ying Shen and
                  Jian Luan and
                  Yang Deng},
  editor       = {Wanxiang Che and
                  Joyce Nabende and
                  Ekaterina Shutova and
                  Mohammad Taher Pilehvar},
  title        = {Browsing Like Human: {A} Multimodal Web Agent with Experiential Fast-and-Slow
                  Thinking},
  booktitle    = {Proceedings of the 63rd Annual Meeting of the Association for Computational
                  Linguistics (Volume 1: Long Papers), {ACL} 2025, Vienna, Austria,
                  July 27 - August 1, 2025},
  pages        = {14232--14251},
  publisher    = {Association for Computational Linguistics},
  year         = {2025},
  bibsource    = {dblp computer science bibliography, https://dblp.org}
}

@inproceedings{MP5,
  author       = {Yiran Qin and
                  Enshen Zhou and
                  Qichang Liu and
                  Zhenfei Yin and
                  Lu Sheng and
                  Ruimao Zhang and
                  Yu Qiao and
                  Jing Shao},
  title        = {{MP5:} {A} Multi-modal Open-ended Embodied System in Minecraft via
                  Active Perception},
  booktitle    = {{IEEE/CVF} Conference on Computer Vision and Pattern Recognition,
                  {CVPR} 2024, Seattle, WA, USA, June 16-22, 2024},
  pages        = {16307--16316},
  publisher    = {{IEEE}},
  year         = {2024},
  doi          = {10.1109/CVPR52733.2024.01543},
  bibsource    = {dblp computer science bibliography, https://dblp.org}
}

@article{JARVIS-1,
  author       = {Zihao Wang and
                  Shaofei Cai and
                  Anji Liu and
                  Yonggang Jin and
                  Jinbing Hou and
                  Bowei Zhang and
                  Haowei Lin and
                  Zhaofeng He and
                  Zilong Zheng and
                  Yaodong Yang and
                  Xiaojian Ma and
                  Yitao Liang},
  title        = {{JARVIS-1:} Open-World Multi-Task Agents With Memory-Augmented Multimodal
                  Language Models},
  journal      = {{IEEE} Trans. Pattern Anal. Mach. Intell.},
  volume       = {47},
  number       = {3},
  pages        = {1894--1907},
  year         = {2025},
  doi          = {10.1109/TPAMI.2024.3511593},
  bibsource    = {dblp computer science bibliography, https://dblp.org}
}

@inproceedings{Emma-Alfworld,
  author       = {Yijun Yang and
                  Tianyi Zhou and
                  Kanxue Li and
                  Dapeng Tao and
                  Lusong Li and
                  Li Shen and
                  Xiaodong He and
                  Jing Jiang and
                  Yuhui Shi},
  title        = {Embodied Multi-Modal Agent trained by an {LLM} from a Parallel TextWorld},
  booktitle    = {{IEEE/CVF} Conference on Computer Vision and Pattern Recognition,
                  {CVPR} 2024, Seattle, WA, USA, June 16-22, 2024},
  pages        = {26265--26275},
  publisher    = {{IEEE}},
  year         = {2024},
  doi          = {10.1109/CVPR52733.2024.02482},
  bibsource    = {dblp computer science bibliography, https://dblp.org}
}

@article{MC-Planner,
  author       = {Zihao Wang and
                  Shaofei Cai and
                  Anji Liu and
                  Xiaojian Ma and
                  Yitao Liang},
  title        = {Describe, Explain, Plan and Select: Interactive Planning with Large
                  Language Models Enables Open-World Multi-Task Agents},
  journal      = {CoRR},
  volume       = {abs/2302.01560},
  year         = {2023},
  doi          = {10.48550/ARXIV.2302.01560},
  eprinttype    = {arXiv},
  eprint       = {2302.01560},
  bibsource    = {dblp computer science bibliography, https://dblp.org}
}

@inproceedings{Octopus,
  author       = {Jingkang Yang and
                  Yuhao Dong and
                  Shuai Liu and
                  Bo Li and
                  Ziyue Wang and
                  Haoran Tan and
                  Chencheng Jiang and
                  Jiamu Kang and
                  Yuanhan Zhang and
                  Kaiyang Zhou and
                  Ziwei Liu},
  editor       = {Ales Leonardis and
                  Elisa Ricci and
                  Stefan Roth and
                  Olga Russakovsky and
                  Torsten Sattler and
                  G{\"{u}}l Varol},
  title        = {Octopus: Embodied Vision-Language Programmer from Environmental Feedback},
  booktitle    = {Computer Vision - {ECCV} 2024 - 18th European Conference, Milan, Italy,
                  September 29-October 4, 2024, Proceedings, Part {I}},
  series       = {Lecture Notes in Computer Science},
  volume       = {15059},
  pages        = {20--38},
  publisher    = {Springer},
  year         = {2024},
  doi          = {10.1007/978-3-031-73232-4\_2},
  bibsource    = {dblp computer science bibliography, https://dblp.org}
}

@inproceedings{STEVE,
  author       = {Zhonghan Zhao and
                  Wenhao Chai and
                  Xuan Wang and
                  Li Boyi and
                  Shengyu Hao and
                  Shidong Cao and
                  Tian Ye and
                  Gaoang Wang},
  editor       = {Ales Leonardis and
                  Elisa Ricci and
                  Stefan Roth and
                  Olga Russakovsky and
                  Torsten Sattler and
                  G{\"{u}}l Varol},
  title        = {See and Think: Embodied Agent in Virtual Environment},
  booktitle    = {Computer Vision - {ECCV} 2024 - 18th European Conference, Milan, Italy,
                  September 29-October 4, 2024, Proceedings, Part {VIII}},
  series       = {Lecture Notes in Computer Science},
  volume       = {15066},
  pages        = {187--204},
  publisher    = {Springer},
  year         = {2024},
  doi          = {10.1007/978-3-031-73242-3\_11},
  bibsource    = {dblp computer science bibliography, https://dblp.org}
}

@inproceedings{LLM-REGRESS,
  author       = {Xiaotian Liu and
                  Ali Pesaranghader and
                  Hanze Li and
                  Punyaphat Sukcharoenchaikul and
                  Jaehong Kim and
                  Tanmana Sadhu and
                  Hyejeong Jeon and
                  Scott Sanner},
  editor       = {Wanxiang Che and
                  Joyce Nabende and
                  Ekaterina Shutova and
                  Mohammad Taher Pilehvar},
  title        = {Open-World Planning via Lifted Regression with LLM-Inferred Affordances
                  for Embodied Agents},
  booktitle    = {Proceedings of the 63rd Annual Meeting of the Association for Computational
                  Linguistics (Volume 1: Long Papers), {ACL} 2025, Vienna, Austria,
                  July 27 - August 1, 2025},
  pages        = {20881--20897},
  publisher    = {Association for Computational Linguistics},
  year         = {2025},
  bibsource    = {dblp computer science bibliography, https://dblp.org}
}

@article{GPT-Driver,
  author       = {Jiageng Mao and
                  Yuxi Qian and
                  Hang Zhao and
                  Yue Wang},
  title        = {GPT-Driver: Learning to Drive with {GPT}},
  journal      = {CoRR},
  volume       = {abs/2310.01415},
  year         = {2023},
  doi          = {10.48550/ARXIV.2310.01415},
  eprinttype    = {arXiv},
  eprint       = {2310.01415},
  bibsource    = {dblp computer science bibliography, https://dblp.org}
}

@inproceedings{DriveLikeAHuman,
  author       = {Daocheng Fu and
                  Xin Li and
                  Licheng Wen and
                  Min Dou and
                  Pinlong Cai and
                  Botian Shi and
                  Yu Qiao},
  title        = {Drive Like a Human: Rethinking Autonomous Driving with Large Language
                  Models},
  booktitle    = {{IEEE/CVF} Winter Conference on Applications of Computer Vision Workshops,
                  {WACVW} 2024 - Workshops, Waikoloa, HI, USA, January 1-6, 2024},
  pages        = {910--919},
  publisher    = {{IEEE}},
  year         = {2024},
  doi          = {10.1109/WACVW60836.2024.00102},
  bibsource    = {dblp computer science bibliography, https://dblp.org}
}

@article{GPT4V-AD-Exploration,
  author       = {Licheng Wen and
                  Xuemeng Yang and
                  Daocheng Fu and
                  Xiaofeng Wang and
                  Pinlong Cai and
                  Xin Li and
                  Tao Ma and
                  Yingxuan Li and
                  Linran Xu and
                  Dengke Shang and
                  Zheng Zhu and
                  Shaoyan Sun and
                  Yeqi Bai and
                  Xinyu Cai and
                  Min Dou and
                  Shuanglu Hu and
                  Botian Shi and
                  Yu Qiao},
  title        = {On the Road with GPT-4V(ision): Early Explorations of Visual-Language
                  Model on Autonomous Driving},
  journal      = {CoRR},
  volume       = {abs/2311.05332},
  year         = {2023},
  doi          = {10.48550/ARXIV.2311.05332},
  eprinttype    = {arXiv},
  eprint       = {2311.05332},
  bibsource    = {dblp computer science bibliography, https://dblp.org}
}

@inproceedings{CityNavAgent,
  author       = {Weichen Zhang and
                  Chen Gao and
                  Shiquan Yu and
                  Ruiying Peng and
                  Baining Zhao and
                  Qian Zhang and
                  Jinqiang Cui and
                  Xinlei Chen and
                  Yong Li},
  editor       = {Wanxiang Che and
                  Joyce Nabende and
                  Ekaterina Shutova and
                  Mohammad Taher Pilehvar},
  title        = {CityNavAgent: Aerial Vision-and-Language Navigation with Hierarchical
                  Semantic Planning and Global Memory},
  booktitle    = {Proceedings of the 63rd Annual Meeting of the Association for Computational
                  Linguistics (Volume 1: Long Papers), {ACL} 2025, Vienna, Austria,
                  July 27 - August 1, 2025},
  pages        = {31292--31309},
  publisher    = {Association for Computational Linguistics},
  year         = {2025},
  bibsource    = {dblp computer science bibliography, https://dblp.org}
}

@inproceedings{RILA,
  author       = {Zeyuan Yang and
                  Jiageng Lin and
                  Peihao Chen and
                  Anoop Cherian and
                  Tim K. Marks and
                  Jonathan Le Roux and
                  Chuang Gan},
  title        = {{RILA:} Reflective and Imaginative Language Agent for Zero-Shot Semantic
                  Audio-Visual Navigation},
  booktitle    = {{IEEE/CVF} Conference on Computer Vision and Pattern Recognition,
                  {CVPR} 2024, Seattle, WA, USA, June 16-22, 2024},
  pages        = {16251--16261},
  publisher    = {{IEEE}},
  year         = {2024},
  doi          = {10.1109/CVPR52733.2024.01538},
  bibsource    = {dblp computer science bibliography, https://dblp.org}
}

@article{LLaVA-Interactive,
  author       = {Wei{-}Ge Chen and
                  Irina Spiridonova and
                  Jianwei Yang and
                  Jianfeng Gao and
                  Chunyuan Li},
  title        = {LLaVA-Interactive: An All-in-One Demo for Image Chat, Segmentation,
                  Generation and Editing},
  journal      = {CoRR},
  volume       = {abs/2311.00571},
  year         = {2023},
  doi          = {10.48550/ARXIV.2311.00571},
  eprinttype    = {arXiv},
  eprint       = {2311.00571},
  bibsource    = {dblp computer science bibliography, https://dblp.org}
}

@article{Visual-ChatGPT,
  author       = {Chenfei Wu and
                  Shengming Yin and
                  Weizhen Qi and
                  Xiaodong Wang and
                  Zecheng Tang and
                  Nan Duan},
  title        = {Visual ChatGPT: Talking, Drawing and Editing with Visual Foundation
                  Models},
  journal      = {CoRR},
  volume       = {abs/2303.04671},
  year         = {2023},
  doi          = {10.48550/ARXIV.2303.04671},
  eprinttype    = {arXiv},
  eprint       = {2303.04671},
  bibsource    = {dblp computer science bibliography, https://dblp.org}
}

@inproceedings{GenArtist,
  author       = {Zhenyu Wang and
                  Aoxue Li and
                  Zhenguo Li and
                  Xihui Liu},
  editor       = {Amir Globersons and
                  Lester Mackey and
                  Danielle Belgrave and
                  Angela Fan and
                  Ulrich Paquet and
                  Jakub M. Tomczak and
                  Cheng Zhang},
  title        = {GenArtist: Multimodal {LLM} as an Agent for Unified Image Generation
                  and Editing},
  booktitle    = {Advances in Neural Information Processing Systems 38: Annual Conference
                  on Neural Information Processing Systems 2024, NeurIPS 2024, Vancouver,
                  BC, Canada, December 10 - 15, 2024},
  year         = {2024},
  bibsource    = {dblp computer science bibliography, https://dblp.org}
}

@article{Loop-Copilot,
  author       = {Yixiao Zhang and
                  Akira Maezawa and
                  Gus Xia and
                  Kazuhiko Yamamoto and
                  Simon Dixon},
  title        = {Loop Copilot: Conducting {AI} Ensembles for Music Generation and Iterative
                  Editing},
  journal      = {CoRR},
  volume       = {abs/2310.12404},
  year         = {2023},
  doi          = {10.48550/ARXIV.2310.12404},
  eprinttype    = {arXiv},
  eprint       = {2310.12404},
  bibsource    = {dblp computer science bibliography, https://dblp.org}
}

@inproceedings{MusicAgent,
  author       = {Dingyao Yu and
                  Kaitao Song and
                  Peiling Lu and
                  Tianyu He and
                  Xu Tan and
                  Wei Ye and
                  Shikun Zhang and
                  Jiang Bian},
  editor       = {Yansong Feng and
                  Els Lefever},
  title        = {MusicAgent: An {AI} Agent for Music Understanding and Generation with
                  Large Language Models},
  booktitle    = {Proceedings of the 2023 Conference on Empirical Methods in Natural
                  Language Processing, {EMNLP} 2023 - System Demonstrations, Singapore,
                  December 6-10, 2023},
  pages        = {246--255},
  publisher    = {Association for Computational Linguistics},
  year         = {2023},
  doi          = {10.18653/V1/2023.EMNLP-DEMO.21},
  bibsource    = {dblp computer science bibliography, https://dblp.org}
}

@article{WavJourney,
  author       = {Xubo Liu and
                  Zhongkai Zhu and
                  Haohe Liu and
                  Yi Yuan and
                  Meng Cui and
                  Qiushi Huang and
                  Jinhua Liang and
                  Yin Cao and
                  Qiuqiang Kong and
                  Mark D. Plumbley and
                  Wenwu Wang},
  title        = {WavJourney: Compositional Audio Creation with Large Language Models},
  journal      = {CoRR},
  volume       = {abs/2307.14335},
  year         = {2023},
  doi          = {10.48550/ARXIV.2307.14335},
  eprinttype    = {arXiv},
  eprint       = {2307.14335},
  bibsource    = {dblp computer science bibliography, https://dblp.org}
}

@inproceedings{empirical_security,
  author       = {Zhuoxiang Shen and 
                  Jiarun Dai and
                  Yuan Zhang and
                  Min Yang},
  title        = {Security Debt in LLM Agent Applications: A Measurement Study of Vulnerabilities andMitigation Trade-offs},
  booktitle    = {40th {IEEE/ACM} International Conference on Automated Software Engineering,
                  {ASE} 2025, Seoul, South Korea, November 16-20, 2025},
  publisher    = {{ACM/IEEE}},
  year         = {2025}
}

@article{empirical_compliance,
  author       = {Ilija Lichkovski and
                  Alexander M{\"{u}}ller and
                  Mariam Ibrahim and
                  Tiwai Mhundwa},
  title        = {EU-Agent-Bench: Measuring Illegal Behavior of {LLM} Agents Under {EU}
                  Law},
  journal      = {CoRR},
  volume       = {abs/2510.21524},
  year         = {2025},
  doi          = {10.48550/ARXIV.2510.21524},
  eprinttype    = {arXiv},
  eprint       = {2510.21524},
  bibsource    = {dblp computer science bibliography, https://dblp.org}
}

@article{agent_fix_agent,
  author       = {Alfin Wijaya Rahardja and
                  Junwei Liu and
                  Weitong Chen and
                  Zhenpeng Chen and
                  Yiling Lou},
  title        = {Can Agents Fix Agent Issues?},
  journal      = {CoRR},
  volume       = {abs/2505.20749},
  year         = {2025},
  doi          = {10.48550/ARXIV.2505.20749},
  eprinttype    = {arXiv},
  eprint       = {2505.20749},
  bibsource    = {dblp computer science bibliography, https://dblp.org}
}

@article{understand_SWEA,
  author       = {Islem Bouzenia and
                  Michael Pradel},
  title        = {Understanding Software Engineering Agents: {A} Study of Thought-Action-Result
                  Trajectories},
  journal      = {CoRR},
  volume       = {abs/2506.18824},
  year         = {2025},
  doi          = {10.48550/ARXIV.2506.18824},
  eprinttype    = {arXiv},
  eprint       = {2506.18824},
  bibsource    = {dblp computer science bibliography, https://dblp.org}
}

@article{understand_SWEA_2,
  author       = {Ira Ceka and
                  Saurabh Pujar and
                  Shyam Ramji and
                  Luca Buratti and
                  Gail E. Kaiser and
                  Baishakhi Ray},
  title        = {Understanding Software Engineering Agents Through the Lens of Traceability:
                  An Empirical Study},
  journal      = {CoRR},
  volume       = {abs/2506.08311},
  year         = {2025},
  doi          = {10.48550/ARXIV.2506.08311},
  eprinttype    = {arXiv},
  eprint       = {2506.08311},
  bibsource    = {dblp computer science bibliography, https://dblp.org}
}

@inproceedings{understand_codeA,
  author       = {Ruofan Lu and
                  Yichen Li and
                  Yintong Huo},
  title        = {Exploring Autonomous Agents: {A} Closer Look at Why They Fail When
                  Completing Tasks},
  booktitle    = {40th {IEEE/ACM} International Conference on Automated Software Engineering,
                  {ASE} 2025, Seoul, Korea, Republic of, November 16-20, 2025},
  pages        = {3856--3860},
  year         = {2025},
  doi          = {10.1109/ASE63991.2025.00330},
  bibsource    = {dblp computer science bibliography, https://dblp.org}
}

@article{llm_agent_module,
  author       = {Kunlun Zhu and
                  Zijia Liu and
                  Bingxuan Li and
                  Muxin Tian and
                  Yingxuan Yang and
                  Jiaxun Zhang and
                  Pengrui Han and
                  Qipeng Xie and
                  Fuyang Cui and
                  Weijia Zhang and
                  Xiaoteng Ma and
                  Xiaodong Yu and
                  Gowtham Ramesh and
                  Jialian Wu and
                  Zicheng Liu and
                  Pan Lu and
                  James Zou and
                  Jiaxuan You},
  title        = {Where {LLM} Agents Fail and How They can Learn From Failures},
  journal      = {CoRR},
  volume       = {abs/2509.25370},
  year         = {2025},
  doi          = {10.48550/ARXIV.2509.25370},
  eprinttype   = {arXiv},
  eprint       = {2509.25370},
  bibsource    = {dblp computer science bibliography, https://dblp.org}
}

@article{multiple_agent,
  author       = {Mert Cemri and
                  Melissa Z. Pan and
                  Shuyi Yang and
                  Lakshya A. Agrawal and
                  Bhavya Chopra and
                  Rishabh Tiwari and
                  Kurt Keutzer and
                  Aditya G. Parameswaran and
                  Dan Klein and
                  Kannan Ramchandran and
                  Matei Zaharia and
                  Joseph E. Gonzalez and
                  Ion Stoica},
  title        = {Why Do Multi-Agent {LLM} Systems Fail?},
  journal      = {CoRR},
  volume       = {abs/2503.13657},
  year         = {2025},
  doi          = {10.48550/ARXIV.2503.13657},
  eprinttype   = {arXiv},
  eprint       = {2503.13657},
  bibsource    = {dblp computer science bibliography, https://dblp.org}
}

@article{platform_orchestrated_agent,
  author       = {Xuyan Ma and
                  Xiaofei Xie and
                  Yawen Wang and
                  Junjie Wang and
                  Boyu Wu and
                  Mingyang Li and
                  Qing Wang},
  title        = {Diagnosing Failure Root Causes in Platform-Orchestrated Agentic Systems:
                  Dataset, Taxonomy, and Benchmark},
  journal      = {CoRR},
  volume       = {abs/2509.23735},
  year         = {2025},
  doi          = {10.48550/ARXIV.2509.23735},
  eprinttype   = {arXiv},
  eprint       = {2509.23735},
  bibsource    = {dblp computer science bibliography, https://dblp.org}
}

@article{understand_SA,
  author       = {Jianshuo Dong and
                  Sheng Guo and
                  Hao Wang and
                  Zhuotao Liu and
                  Tianwei Zhang and
                  Ke Xu and
                  Minlie Huang and
                  Han Qiu},
  title        = {SafeSearch: Automated Red-Teaming for the Safety of LLM-Based Search
                  Agents},
  journal      = {CoRR},
  volume       = {abs/2509.23694},
  year         = {2025},
  doi          = {10.48550/ARXIV.2509.23694},
  eprinttype    = {arXiv},
  eprint       = {2509.23694},
  bibsource    = {dblp computer science bibliography, https://dblp.org}
}

@inproceedings{multi-modal-hallucination,
  author       = {Yexing Du and
                  Kaiyuan Liu and
                  Youcheng Pan and
                  Zheng Chu and
                  Bo Yang and
                  Xiaocheng Feng and
                  Ming Liu and
                  Yang Xiang},
  title        = {{CCFQA:} {A} Benchmark for Cross-Lingual and Cross-Modal Speech and
                  Text Factuality Evaluation},
  booktitle    = {Fortieth {AAAI} Conference on Artificial Intelligence, Thirty-Eighth
                  Conference on Innovative Applications of Artificial Intelligence,
                  Sixteenth Symposium on Educational Advances in Artificial Intelligence,
                  {AAAI} 2026, Singapore, January 20-27, 2026},
  pages        = {30575--30583},
  year         = {2026},
  doi          = {10.1609/AAAI.V40I36.40312},
  bibsource    = {dblp computer science bibliography, https://dblp.org}
}
